\documentclass[12pt,draftclsnofoot,onecolumn]{IEEEtran}

\usepackage{amssymb}
\usepackage{amsmath}
\usepackage{amsmath,bm}
\usepackage{amsthm}
\usepackage{amsfonts}
\usepackage{graphicx}
\usepackage{subfigure}
\usepackage{cite}
\usepackage{enumerate}
\usepackage{url}
\usepackage[noend]{algpseudocode}
\usepackage{algorithmicx,algorithm}
\usepackage{mathtools}
\usepackage{algpseudocode}
\usepackage{color, soul}
\usepackage{exscale}
\usepackage{relsize}
\allowdisplaybreaks[4]

\begin{document}
	
\newtheorem{definition}{\bf ~~Definition}
\newtheorem{observation}{\bf ~~Observation}
\newtheorem{theorem}{\bf ~~Theorem}
\newtheorem{proposition}{\bf ~~Proposition}
\newtheorem{remark}{\bf ~~Remark}

\renewcommand{\algorithmicrequire}{\textbf{Input:}} % Use Input in the format of Algorithm
\renewcommand{\algorithmicensure}{\textbf{Output:}} % Use Output in the format of Algorithm

\title{\LARGE{Hybrid Beamforming for Reconfigurable Intelligent Surface based Multi-user Communications: Achievable Rates with Limited Discrete Phase Shifts}}
\author{\small
	\IEEEauthorblockN{
		{Boya Di}, \IEEEmembership{\small{Member, IEEE}},
		{Hongliang Zhang}, \IEEEmembership{\small{Member, IEEE}},
		{Lingyang Song}, \IEEEmembership{\small{Fellow, IEEE}},\\
		{Yonghui Li}, \IEEEmembership{\small{Fellow, IEEE}},
		{Zhu Han}, \IEEEmembership{\small{Fellow, IEEE}},
		{and H. Vincent Poor}, \IEEEmembership{\small{Fellow, IEEE}}
	}\\
	
	\thanks{\scriptsize{Boya Di is with Department of Electronics Engineering, Peking University, Beijing, China and with Department of Computing, Imperial College London, London, UK (email: diboya92@gmail.com).}}
	\thanks{\scriptsize{Hongliang Zhang is with Department of Electronics Engineering, Peking University, Beijing, China and with Department of Electronical and Computer Engineering, University of Houston, Houston, TX, USA (email: diboya92@gmail.com).}}
	\thanks{\scriptsize{Lingyang Song is with Department of Electronics Engineering, Peking University, Beijing, China (email: lingyang.song@pku.edu.cn).}}
	\thanks{\scriptsize{Yonghui Li is with School of Electrical and Information Engineering, the University of Sydney, Australia (yonghui.li@sydney.edu.au).}}
	\thanks{\scriptsize{Zhu Han is with Department of Electronical and Computer Engineering, University of Houston, Houston, TX, USA (zhan2@uh.edu).}}
	\thanks{\scriptsize{H. Vincent Poor is with School of Engineering and Applied Science, Princeton University, NY, USA (poor@princeton.edu).}} }
\maketitle
\vspace{-1.2cm}
\begin{abstract}
Reconfigurable intelligent surface (RIS) has drawn considerable attention from the research society recently, which creates favorable propagation conditions by controlling the phase shifts of the reflected waves at the surface, thereby enhancing wireless transmissions. In this paper, we study a downlink multi-user system where the transmission from a multi-antenna base station (BS) to various users is achieved by the RIS reflecting the incident signals of the BS towards the users. Unlike most existing works, we consider the practical case where only a limited number of discrete phase shifts can be realized by the finite-sized RIS. Based on the reflection-dominated one-hop propagation model between the BS and users via the RIS, a hybrid beamforming scheme is proposed and the sum-rate maximization problem is formulated. Specifically, the continuous digital beamforming and discrete RIS-based analog beamforming are performed at the BS and the RIS, respectively, and an iterative algorithm is designed to solve this problem. Both theoretical analysis and numerical validations show that the RIS-based system can achieve a good sum-rate performance by setting a reasonable size of RIS and a small number of discrete phase shifts.
\end{abstract}


\newpage

\section{Introduction}
%第一段： 当前通信的瓶颈：通信质量受环境影响大，现有技术难以解决
%现在用户需求很高，而且通信环境通常不稳定。传统的方法比如relay和MIMO提升了很多，但是以。。。为代价。

The past decade has witnessed an enormous increase in the number of mobile devices~\cite{EWP-2011}, triggering urgent needs for high-speed and seamless data services in future wireless systems. To meet such demands, one fundamental issue is how to improve the link quality in the complicated time-varying wireless environments involving unpredictable fading and strong shadowing effects. Various technologies have been developed such as relay~\cite{L-2011} and massive multiple input and multiple output (MIMO)~\cite{EOFT-2014}, aiming to actively strengthen the target signals by forwarding and taking advantage of multi-path effects, respectively. However, these techniques require extra hardware implementation with inevitable power consumption and high complexity for signal processing, and the quality of services is also not always guaranteed in harsh propagation environments.
%
%第二段： fortunately, owing to the development of meta-materials, 一种新的。。。可用于改善。。。

Recently, the development of meta-surfaces~\cite{DOLPN-2017} has given rise to a new transmission technique named reconfigurable intelligent surface (RIS), which shapes the propagation environment into a desirable form by controlling the electromagnetic response of multiple scatters~\cite{MMDAMCVGJHJAGM-2019}. Specifically, the RIS is an ultra-thin surface inlaid with multiple sub-wavelength scatters, i.e., RIS elements, whose electromagnetic response (such as phase shifts) can be controlled by simple programmable PIN diodes~\cite{TDR-2010}. Based on the ON/OFF functions of PIN diodes, only a limited number of discrete phases shifts can be achieved by the RIS~\cite{LHCYYACT-2019}. Instead of scattered waves emanated from traditional antennas, the sub-wavelength separation between adjacent RIS elements enables the refracted and reflected waves to be generated via superposition of incident waves at the surface~\cite{HAN-2016}. Benefited from such a programmable characteristic of molding the wavefronts into desired shapes, the RIS serves as a part of reconfigurable propagation environment such that the received signals are directly reflected towards the receivers without any extra cost of power sources or hardwares~\cite{NOYIMD-2019}, thereby improving the link quality and coverage.

To exploit the potential of RIS techniques, many existing works have considered the RIS as a reflection-type surface deployed between sources and destinations in either point-to-point communications~\cite{MMDAMCVGJHJAGM-2019,EMJMMR-2019,XDR-2019,YWSCX-2019} or multi-user systems~\cite{CGAMC-2018,SJI-2019,QAAMM-2019,QR-2019} for higher data rates or energy saving. In~\cite{XDR-2019}, a point-to-point RIS-assisted multi-input single-out system has been investigated where the beamformer at the transmitter and continuous phase shifts of the RIS are jointly optimized to maximize the sum rate.  In~\cite{QAAMM-2019}, a channel estimation protocol has been proposed for a multi-user RIS-assisted system and the continuous phase shifts have been designed to maximize the minimum user date. In~\cite{QR-2019}, the authors have minimized the transmit power of the access point by optimizing the continuous digital beamforming and discrete phase shifts. An algorithm has been designed for the single-user case and extended to the multi-user case.

However, two major issues still remain to be further discussed in the open literature.
\begin{itemize}
	\item \emph{For the multi-user case, how to determine the limited discrete phase shifts directly such that the inter-user interference can be eliminated? How does the quantization level influence the sum rate of the system?}
	\item \emph{Considering the strengthened coupling between propagation and discrete phase shifts brought by the dominant reflection ray via the RIS, how do we design the size of RIS and perform beamforming in a multi-antenna system to achieve the maximum sum rate? }
\end{itemize}
%       介绍一下RIS的原理 （可以参照system model，还是要说一下信道的）
%
%第三段： We consider a downlink ... system ... However, ...(即引出hybrid beamforming的动机)
%拆成两段，一段引出信道模型的不同，一段给出hybrid beamforming的动机（based on channel）

To address the above issues, in this paper, we consider a downlink multi-user multi-antenna system where the direct links between the multi-antenna base station (BS) and users suffer from deep shadowing. To provide high-quality data services, a RIS with limited discrete phase shifts is deployed between the BS and users such that the signals sent by the BS are reflected by the RIS towards the users. Since the incident waves are reflected rather than scattered at the RIS, it is the reflection-based one-hop ray~\cite{G-2005} via the RIS that dominates the propagation between the BS and users. Therefore, the propagation model in the RIS-based system differs from those for traditional two-hop relays and one-hop direct links in MIMO systems, revealing an inner connection between the phase shifts and the propagation paths.

%forming different configuration patterns. Moreover,

%考虑实际条件，相位只能离散调整。由于反射波的存在，传播模型是一个以反射径为主径的模型，这与传统的relay 两跳模型和MIMO直射径模型不同。并且导致了传播相位的变化。

%To achieve ..., it is very important to ...选择。。。并且对信号进行预编码非常重要。configuration pattern selection实际上实现了analog beamforming。However, 由于反射信道模型，a new hybrid beamforming scheme is required to depict...
To achieve better directional reflection rays towards the desired users, it is vitally important to determine the phase shifts of all RIS elements, the process of which is also known as the \emph{RIS configuration}. Such built-in programmable configuration~\cite{MMDAMCVGJHJAGM-2019} is actually equivalent to analog beamforming, realized by the RIS inherently. Since the RIS elements do not have any digital processing capability, we consider the hybrid beamforming (HBF)~\cite{FW-2016} consisting of the digital beamforming at the BS and RIS configuration based analog beamforming. A novel HBF scheme for RIS-based communications with discrete phase shifts is thus required for better shaping the propagation environment and sum rate maximization.

Designing an HBF scheme presents several major challenges. \emph{First}, the reflection-dominated one-hop propagation and the RIS configuration based analog beamforming are coupled with each other, rendering the optimal scheme very hard to be obtained. The traditional beamforming schemes with separate channel matrix and analog beamformer do not work any more. \emph{Second}, discrete phase shifts required by the RIS renders the sum rate maximization to be a mixed integer programming problem, which is non-trivial to be solved especially in the complex domain. \emph{Third}, given the dense placing of the RIS elements, the correlation between elements may degrade the data rate performance. Thus, it is necessary to explore how the achievable rate is influenced by the size of RIS, which is challenging due to the complicated propagation environments especially in a multi-user case.

%秩的事：如果只有一个用户，是没有影响的（所以多用户场景更值得分析）

Through solving the above challenges, we aim to design an HBF scheme for the RIS-based multi-user system with limited discrete phase shifts to maximize the sum rate. Our main contributions can be summarized below.
%1. 建了新的信道
%2. 提出了新的hybrid beamforming scheme，分析了。。。，
%3. 设计了算法并进行了仿真
\begin{itemize}
	\item We consider a downlink RIS-based multi-user system where a RIS with limited discrete phase shifts reflects signals from the BS towards various users. Given a reflection-based one-hop propagation model, we design an HBF scheme where the digital beamforming is performed at the BS, and the RIS-based analog beamforming is conducted at the RIS.
	\item A mixed-integer sum rate maximization problem for RIS-based HBF is formulated and decomposed into two subproblems. We propose an iterative algorithm in which the digital beamforming subproblem is solved by zero-forcing (ZF) beamforming with power allocation and the RIS-based analog beamforming is solved by the outer approximation.
	\item We prove that the proposed RIS-based HBF scheme can save as much as half of the radio frequency (RF) chains compared to traditional HBF schemes. Extending from our theoretical analysis on the pure Line-of-Sight (LoS) case, we reveal the influence of the size of RIS and the number of discrete phase shifts on the sum rate both theoretically and numerically.
\end{itemize}

The rest of this paper is organized as follows. In Section II, we introduce the system model of the downlink RIS-based MU multi-antenna system. The frequency-response model of the RIS and the channel model are derived. In Seciton III, the HBF scheme for the RIS-based system is proposed. A sum rate maximization problem is formulated and decomposed into two subproblems: digital beamforming and RIS configuration based analog beamforming. An iterative algorithm is developed in Section IV to solve the above two subproblems and a sub-optimal solution is obtained. In Section V, we compare the RIS-based HBF scheme with the traditional one theoretically, and discuss how to achieve the maximum sum rate in the pure LoS case. The complexity and convergence of the proposed algorithm are also analyzed. Numerical results in Section VI evaluate the performance of the proposed algorithm and validate our analysis. Finally, conclusions are drawn in Section VII.

\emph{Notations:} Scalars are denoted by italic letters, vectors and matrices are denoted by bold-face lower-case and uppercase letters, respectively. For a complex-valued vector $\bm{x}$, $\|\bm{x}\|$ denotes its Euclidean norm, and $\text{diag}(\bm{x})$ denotes a diagonal matrix whose diagonal element is the corresponding element in $\bm{x}$. For a square matrix $\bm{S}$, $\text{Tr}(\bm{S})$ denotes its trace. For any general matrix $\bm{M}$, $\bm{M}^{H}$ and $\bm{M}^T$ denote its conjugate transpose and transpose, respectively. $\bm{I}$, $\bm{0}$, and $\bm{1}$ denote an identity matrix, all-zero and all-one vectors, respectively.

%第四段： 新的挑战 （可能会有different from relay 和 mimo beamforming 这种话）
%
%第五段： existing works
%
%第六段： contributions
%
%第七段： The rest of this paper is organized as follows.
%
%第八段： notions （就是说明一下矩阵用大写粗体，向量用小写粗体，共轭转置的符号啥啥的）

\section{System Model}
In this section, we first introduce the RIS-based downlink multi-user multi-antenna system in which the BS with multiple antennas serves various single-antenna users via the RIS such that the propagation environment can be pre-designed and configured to optimize the system performance. The discrete phase shift model of RIS and the channel model are then constructed, respectively.

\subsection{Scenario Description}
\begin{figure}[!t]
	\centering
	\includegraphics[width=6in]{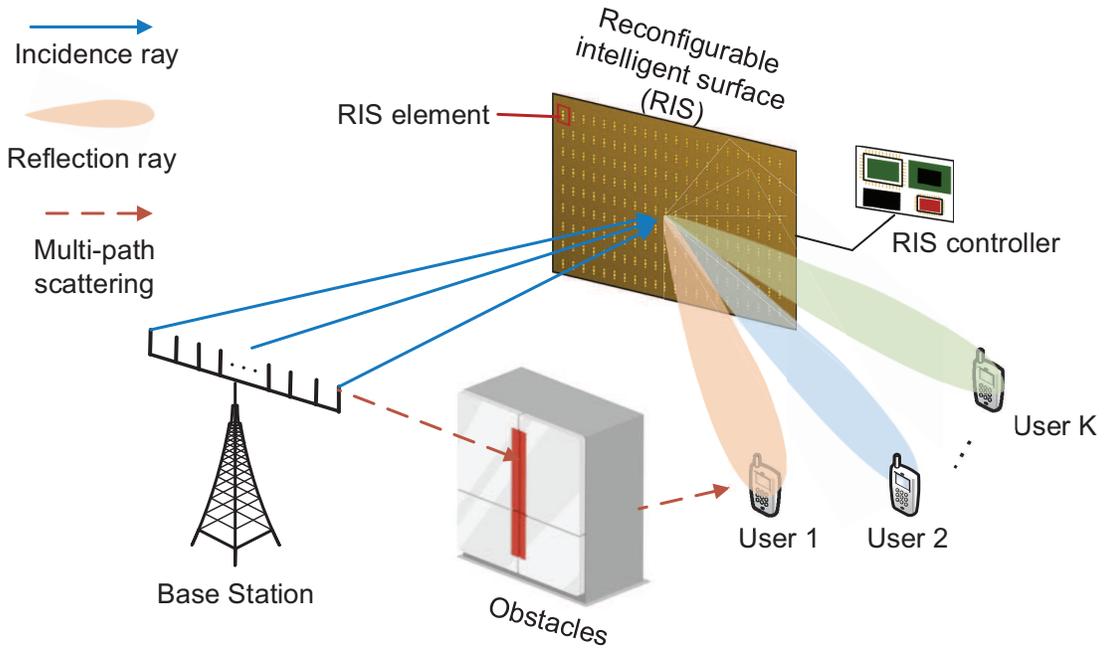}
	%\vspace{-0.5cm}
	\caption{System model of the RIS-based downlink multi-user communication system} \label{system_model}
	%\vspace{-0.9cm}
\end{figure}

Consider a downlink multi-user communication system as shown in Fig.~\ref{system_model} where a BS equipped with $N_t$ antennas transmits to $K$ single-antenna users. Due to the complicated and dynamic wireless environment involving unexpected fading and potential obstacles, the BS - user link may not be stable enough or even be in outage. To alleviate this issue, we consider to deploy an RIS between the BS and users, which reflects the signals from the BS and directly projects to the users by actively shaping the propagation environment into a desirable form.

An RIS consists of $N_R \times N_R$ electrically controlled RIS elements as shown in Fig.~\ref{system_model}, each of which is a sub-wavelength meta-material particle with very small features. An RIS controller, equipped with several PIN diodes, can control the ON/OFF state of the connection between adjacent metal plates where the RIS elements are laid, thereby manipulating the electromagnetic response of RIS elements towards incident waves. Due to these built-in programmable elements, the RIS requires no extra active power sources nor do not have any signal processing capability such as decoding~\cite{MMDAMCVGJHJAGM-2019}. In other words, it serves as a low-cost reconfigurable phased array that only relies on the combination of multiple programmable radiating elements to realize a desired transformation on the transmitted, received, or reflected waves~\cite{NOYIMD-2019}.

%To be specific, we adopt the electrically modulated 2-bit reprogrammable coding meta-surface panel in [4]. The meta-surface panel is composed of a two-dimensional array of electrically controllable meta-surface units, as shown in Fig. 2. For a meta-surface unit, it has three PIN diodes which control the ON and OFF of the connection between adjacent metal plates. By controlling the voltages on the three PIN diodes, the phase characteristics of the reflected radio wave can be modified. To be specific, for the 2-bit coding meta-surface unit, the reflected radio wave can have four types of phase shifts, depending on the on-off conditions of the PIN diodes

%Different from the traditional %number of RF chain blah blah blah (??RF chain?ADC,?????????)

\subsection{Reconfigurable Intelligent Surface with Limited Discrete Phase Shifts}
\begin{figure}[!t]
	\centering
	\includegraphics[width=4in]{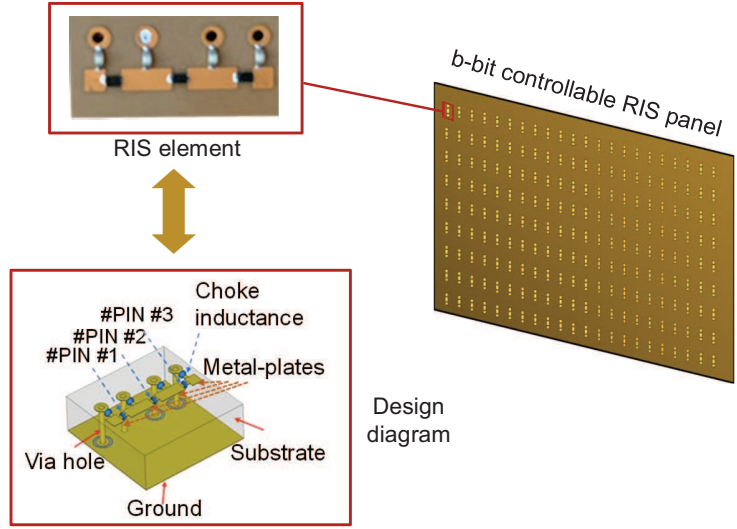}
	%\vspace{-0.5cm}
	\caption{Schematic structure of $b$-bit encoded RIS} \label{element}
	%\vspace{-0.9cm}
\end{figure}

As shown in Fig.~\ref{element}, the RIS is achieved by the $b$-bit re-programmable meta-material, which has been implemented as a set of radiative elements layered on a guiding structure following the wave guide techniques, forming a 2-dimensional (2D) planar antenna array~\cite{LHCYYACT-2019}. Being a miniature radiative element, the field radiated from the RIS element has a \emph{phase} and \emph{amplitude} determined by this element's polarizability, which can be tuned by the RIS controller via multiple PIN diodes (ON/OFF)~\cite{MMDAMCVGJHJAGM-2019}. However, the phase and amplitude introduced by an RIS element are not generated randomly; instead they are constrained by the Lorentzian resonance response~\cite{DOLPN-2017}, which greatly limits the range of phase values. Based on such constraints, one common manner of implementation is to constrain the amplitude and sample the phase values from the finite feasible set~\cite{DOLPN-2017} such that the voltage-controlled diodes can easily manipulate a \emph{discrete} set of phase values at a very low cost.

Specifically, we assume that each RIS element is encoded by the controller (e.g., via PIN diodes) to conduct $2^b$ possible phase shifts to reflect the radio wave. Due to the frequency-selective nature of the meta-materials, these elements only vibrate in resonance with the incoming waves over a narrow band centering at the resonance frequency. Without loss of generality, we denote the frequency response of each element $(l_1, l_2)$ at the $l_1$-th row and $l_2$-th column of the 2D RIS within the considered frequency range as $q_{l_1, l_2}$, $0 \le l_1, l_2 \le N_R - 1$. Since the RIS is $b$-bit controllable, $2^b$ possible configuration modes (i.e., phases) of each $q_{l_1, l_2}$ can be defined according to the Lorentzian resonance response~\cite{DOLPN-2017}.
\begin{equation} \label{q_value}
q_{l_1, l_2} = \frac{j + e^{j\theta_{{l_1},{l_2}}}}{2}, {\theta _{{l_1},{l_2}}} = \frac{{m_{l_1,l_2}\pi }}{{{2^{b - 1}}}},m_{l_1,l_2} \in\left\lbrace 0,1, \ldots ,{2^b} - 1\right\rbrace, 0 \le l_1, l_2 \le N_R - 1,
\end{equation}
where $\theta_{{l}_1,{l}_2}$ denotes the phase shift of RIS element $(l_1, l_2)$. For convenience, we refer to $b$ as the number of \emph{quantization bits}.

%\begin{figure}[!t]
%	\centering
%	\subfigure[]{
%		\label{element} %% label for first subfigure
%		\includegraphics[width=3.1in]{element.eps}}
%	\hspace{0.5cm}
%	\subfigure[]{
%		\label{angle} %% label for second subfigure
%		\includegraphics[width=2.4in]{angle_model.eps}}
%	\vspace{-0.1cm}
%	\caption{(a) Physical structure of 2-bit RIS element; (b) Placement of the antenna arrays at the BS and the RIS.} \label{doublelast} %% label for entire figure
%\end{figure}

Since the 2D RIS is constructed based on the ultra-thin wave guides, the propagation inside the 2D RIS is influenced by both the location and the wave-number of each meta-surface element. The latter one is also a function of the frequency, which reflects the frequency-selective nature of RIS. Denote the position of the element $(l_1, l_2)$ as $p_{l_1, l_2}$. The propagation introduced by this element can then be given by ${e^{ - j{\beta _{{l_1},{l_2}}}\left( \lambda  \right)\frac{{2\pi }}{\lambda }p_{l_1, l_2}}}$, where ${\beta _{{l_1},{l_2}}}\left( \lambda  \right)$ is the wave-number of RIS element $(l_1, l_2)$, and $\lambda$ is the corresponding wave length. For simplicity, we assume that the wave-number ${\beta _{{l_1},{l_2}}}\left( \lambda  \right)$ remains the same for all RIS element within the considered narrow band\footnote{This model can be easily extended to a frequency-selective case where ${\beta _{{l_1},{l_2}}}\left( \lambda  \right)$ varies with the working frequency. The propagation can then be modelled by a filter with finite impulse response~\cite{NOYIMD-2019}.}.

\subsection{Reflection-dominated Channel Model} \label{channel_model}
In this subsection, we model the channel between antenna $0 \le n \le N_t - 1$ of the BS and user $k$. Specifically, instead of the traditional two-hop channel for relays~\cite{EMJMMR-2019}, we use the one-hop reflection ray to model the dominant channel between the BS and users via the RIS which only passively reflects the received signals. The key reason can be detailed below. Due to small spacing between adjacent RIS elements (usually much less than the wavelength), the signals projected onto the surface are no longer just scattered randomly into the open space like those signals spread by the traditional antennas. Instead, the superposition of spherical waves facilitated by a number of miniature scatters enables refracted and reflected waves~\cite{HAN-2016} without any extra decoding or signal forwarding procedures. Therefore, unlike the traditional scattering-based propagation model where signals travel independently along the BS - RIS and RIS - user paths, in the considered scenario signals are only passively reflected by the RIS along the reflection-based path due to the coupling effect of RIS elements.

Moreover, benefited from the directional reflections of the RIS, the BS - RIS - user link is usually stronger than other multipaths as well as the degraded direct link between the BS and the user~\cite{G-2005}. Therefore, we model the channel between the BS and each user $k$ as a Ricean model such that the BS - RIS - user link acts as the dominant ``LoS" component and all the other paths together form the ``non-LoS (NLOS)" component.

%Therefore, we model the channel between the BS and each user $k$ as a Ricean model such that the BS - RIS - user link acts as the dominant ``line-of-sight (LoS)" component and all the other paths together form the ``non-LoS (NLOS)" component.

Specifically, let $D_{l_1, l_2}^{\left(n\right)}$ and $d_{l_1, l_2}^{\left(k\right)}$ denote the distance between antenna $n$ and RIS element $\left( l_1, l_2 \right)$, and that between user $k$ and RIS element $\left( l_1, l_2 \right)$, respectively. The ``LoS" channel between the signal transmitted by the BS at antenna $1 \le n \le N_t$ to user $k$ via RIS element $\left( l_1, l_2 \right)$ can be given by
\begin{equation}
h_{{l_1},{l_2}}^{\left( {k,n} \right)} = {\left[ {D_{{l_1},{l_2}}^{\left( n \right)} + d_{{l_1},{l_2}}^{\left( k \right)}} \right]^{ - \alpha }} \cdot {e^{ - j\beta_{{l_1},{l_2}}\left(\lambda \right) \frac{{2\pi }}{\lambda }\left[ {D_{{l_1},{l_2}}^{\left( n \right)} + d_{{l_1},{l_2}}^{\left( k \right)}} \right]}},
\end{equation}
where $\alpha$ is the path loss parameter. Therefore, the channel model between each antenna $n$ of the BS and user $k$ via RIS element $\left(l_1, l_2 \right)$ can be written by
\begin{equation}
\tilde h_{{l_1},{l_2}}^{\left( {k,n} \right)} = \sqrt {\frac{\kappa }{{1 + \kappa }}} h_{{l_1},{l_2}}^{\left( {k,n} \right)} + \sqrt {\frac{1}{{1 + \kappa }}} PL\left( {D_{{l_1},{l_2}}^{\left( n \right)} + d_{{l_1},{l_2}}^{\left( k \right)}} \right)h_{NLOS,({l_1},{l_2})}^{\left( {k,n} \right)},
\end{equation}
where $\kappa$ is the Rician factor, $PL\left( {\cdot} \right)$ is the path loss model for NLOS transmissions, and $h_{NLOS,({l_1},{l_2})}^{\left( {k,n} \right)} \sim {\cal{C}\cal{N}}{\left(0,1\right)}$ is the small-scale NLOS component. Here we assume that the perfect channel state information is known to the BS via communicating with the RIS controller over a dedicated wireless link. A number of channel estimation methods can be found in~\cite{QAAMM-2019,QR-2019}, which is out of the scope of this paper. For the case where channel information is partially known to the BS, we will consider the pure LoS transmission and discuss it in detail in Section V.

\begin{figure}[!t]
	\centering
	\includegraphics[width=5in]{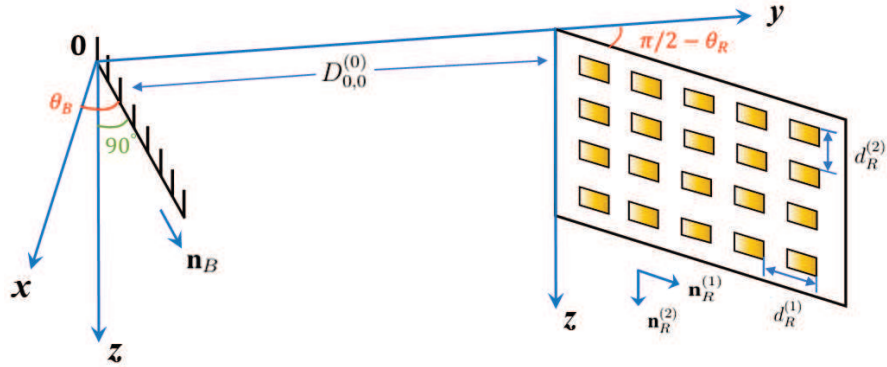}
	%\vspace{-0.5cm}
	\caption{Placement of the antenna arrays at the BS and the RIS} \label{angle}
	%\vspace{-0.9cm}
\end{figure}

\emph{Geometric Model of the Propagation}: We now derive the above distances $D_{l_1, l_2}^{\left(n\right)}$ and $d_{l_1, l_2}^{\left(k\right)}$ by using the geometry information~\cite{FPG-2007}. Without loss of generality, we assume that the uniform planar arrays (UPA) and uniform linear arrays (ULA) are deployed at the RIS and the BS, respectively. The UPA is placed in a way that the surface is perpendicular to the ground. To describe the geometry of the above mentioned UPA and ULA, we employ the spherical coordinates as shown in Fig.~\ref{angle}. For the RIS, we define the local origin as the higher corner of the surface; while for the BS, its local origin is set as one end of the linear array. The $y$ axis is set along the direction of BS antenna 0 - element $\left(0,0\right)$ link, and the $z$ axis is perpendicular to the ground. For convenience, the directions of the ULA and RIS on the $x$-$y$ plane can be depicted by the angles $\theta_B \in \left[0, 2\pi \right]$ and $\theta_R \in \left[0, 2\pi \right]$, respectively, as shown in Fig.~\ref{angle}. We can then define the principle directions of the linear antenna arrays at the BS and the RIS, denoted by ${\textbf{n}}_B$ and $\left\lbrace  {\textbf{n}}_R^{\left(1 \right)}, {\textbf{n}}_R^{\left(2 \right)} \right\rbrace $, as
\begin{subequations} \label{direction}
\begin{align}
& \textbf{n}_B = {\textbf{n}_x}\cos{\theta_B} + {\textbf{n}_y}\sin{\theta_B},\\
& \textbf{n}_R^{\left(1 \right)} = {\textbf{n}_x}\cos{\theta_R} + {\textbf{n}_y}\sin{\theta_R},\\
&\textbf{n}_R^{\left(2 \right)} = \textbf{n}_z,
\end{align}
\end{subequations}
where ${\textbf{n}_x}$, ${\textbf{n}_y}$, and ${\textbf{n}_z}$ are directions of the $x$, $y$, and $z$ axis, respectively.

We denote the uniform separation between any two adjacent elements in the above defined ULA and UPA as $d_{B}$ and $\left\lbrace d_{R}^{\left(1 \right) }, d_{R}^{\left(2 \right) }\right\rbrace $, respectively. The position of any antenna $n$ at the BS can be represented by
\begin{equation} \label{position_BS}
{\textbf{c}}_B^{\left(n\right)} = n{d_B}{\textbf{n}_B},
\end{equation}
and the position of any RIS element $\left( l_1, l_2 \right)$ can be given by
\begin{equation} \label{position_RIS}
{\textbf{c}}_R^{\left(l_1, l_2\right)} = {l_1}d_{R}^{\left(1 \right) }{\textbf{n}_R^{\left(1 \right)}} + {l_2}d_{R}^{\left(2 \right) }{\textbf{n}_R^{\left(2 \right)}} + {D_{0,0}^{\left( 0\right) }}{\textbf{n}_y}.
\end{equation}
Based on $\left( \ref{direction}\right) $ - $\left( \ref{position_RIS}\right) $, the distance between antenna $n$ and RIS element $\left( l_1, l_2 \right)$, i.e., $D_{l_1, l_2}^{\left(n \right) }$, can be calculated as
\begin{subequations} \label{distance_BS_RIS}
\begin{align}
D_{l_1, l_2}^{\left(n \right)} & = \left\| {{\textbf{c}}_R^{\left(l_1, l_2\right)} - {\textbf{c}}_B^{\left(n\right)}} \right\|_2\\
& = \left[ {\left( l_1 d_R^{\left(1\right)} \cos{\theta_R} - n {d_B} \cos{\theta_B}\right)}^2 + {\left( l_1 d_R^{\left(1\right)} \sin{\theta_R} + D_{0,0}^{\left(0\right)} - n {d_B} \sin{\theta_B}\right)}^2 + {\left( l_2 d_R^{\left(2\right)}\right)}^2\right]^{\frac{1}{2}}\\
& \approx \left( l_1 d_R^{\left(1\right)} \sin{\theta_R} + D_{0,0}^{\left(0\right)} - n {d_B} \sin{\theta_B}\right) + \frac{{\left( l_1 d_R^{\left(1\right)} \cos{\theta_R} - n {d_B} \cos{\theta_B}\right)}^2+{\left( l_2 d_R^{\left(2\right)}\right)}^2}{2D_{0,0}^{0}}, \label{approx}
\end{align}
\end{subequations}
where $\left( \ref{approx} \right)$ is obtained by adopting $\sqrt{1 + a} \approx 1 + {a}/{2}$ when $a \ll 1$. Denote the position of each user $k$ as ${\textbf{c}}_k = \left( d_{x,k}, d_{y,k}, d_{z,k} \right)$. The distance between the RIS element and any user $k$ can be expressed by
\begin{equation} \label{distance_RIS_user}
d_{l_1, l_2}^{\left(k \right)} = \left\| {{\textbf{c}}_R^{\left(l_1, l_2\right)} - \textbf{c}}_k \right\|_2.
\end{equation}

Since the distance between any two antennas or RIS elements is much smaller than the distance between the BS and the user, i.e., $D_{l_1, l_2}^{\left(n \right)} + d_{l_1, l_2}^{\left(k \right)} \gg d_R^{\left(1\right)}, d_R^{\left(2\right)}, d_B$, we assume that the path loss of each BS-user link is the same, ignoring the influence brought by different antennas or RIS elements. Therefore, the channel propagation $h_{{l_1},{l_2}}^{\left( {k,n} \right)}$ can be rewritten as
\begin{equation}
h_{{l_1},{l_2}}^{\left( {k,n} \right)} = {\left[ {D_{0,0}^{\left( 0 \right)} + d_{0,0}^{\left( k \right)}} \right]^{ - \alpha }}{e^{ - j\beta_{{l_1},{l_2}}\left(\lambda \right)\frac{{2\pi }}{\lambda }\left[ {D_{l_1,l_2}^{\left( n \right)} + d_{l_1,l_2}^{\left( k \right)}} \right]}},
\end{equation}
where $D_{l_1,l_2}^{\left( n \right)}$ and $d_{l_1,l_2}^{\left( k \right)}$ are given above in $\left(\ref{distance_BS_RIS} \right)$ and $\left( \ref{distance_RIS_user} \right)$.

Based on the propagation characteristics introduced above, we will investigate how RIS can be utilized to assist multi-user transmissions in the following section.

\section{RIS-based Hybrid Beamforming and Problem Formulation for Multi-user Communications}

Note that the RIS usually consists of a large number of RIS elements, which can be viewed as antenna elements fay away from the BS, inherently capable of realizing analog beamforming via RIS configuration. However, these RIS elements do not have any digital processing capacity, requiring signal processing to be carried out at the BS.

\begin{figure}[!t]
	\centering
	\includegraphics[width=6.5in]{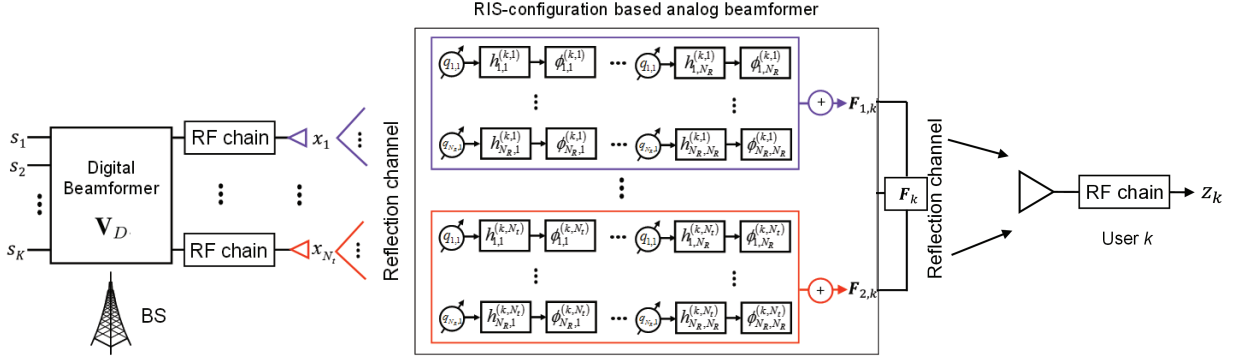}
	%\vspace{-0.5cm}
	\caption{Block diagram of the RIS-based transmission between the BS and user $k$.} \label{transmission}
	%\vspace{-0.9cm}
\end{figure}

In this section, to realize reflected waves towards preferable directions, we present an HBF scheme for RIS-based multi-user communications given the phase shift model and the channel model of RIS-based transmissions in Section II. As shown in Fig.~\ref{transmission}, the digital beamforming is performed at the BS while the analog beamforming is achieved by the RIS with discrete phase shifts. Based on the considered HBF scheme, we formulate a sum rate maximization problem, and then decompose it into the digital beamforming subproblem and the RIS configuration based analog beamforming subproblem.

% in order to approach arbitrarily shaped waves. These RIS elements act as a number of antennas far away from the BS, being part of the propagation environment. However, the RIS elements do not have any digital processing capacity. Therefore, it is not practical to equip each RIS element with one radio frequency (RF) chain and ADC due to the physical isolation between the RIS and the BS as well as the inevitably high cost of fully digital beamforming.

\subsection{Hybrid Beamforming Scheme}
\subsubsection{Digital Beamforming at the BS}

The BS first encodes $K$ different data steams via a digital beamformer, ${\textbf{V}}_D$, of size $N_t \times K$, satisfying $N_t \ge K$. After up-converting the encoded signal over the carrier frequency and allocating the transmit powers, the BS sends users' signals directly through $N_t$ antennas. Denote the intended signal vector for $K$ users as $\emph{\textbf{s}} \in {\mathbb{C}}^{K \times 1}$. The transmitted signals of the BS can be given by
\begin{equation} \label{x_transmit}
\emph{\textbf{x}} = {{\textbf{V}}_D}\emph{\textbf{s}}.
\end{equation}
\subsubsection{RIS Configuration based Analog Beamforming}
After travelling through the reflection-dominated channel introduced in Section~\ref{channel_model}, the received signal at the antenna of user $k$ can then be expressed as
\begin{equation}\label{scalar_signal}
{z_k} = \sum\limits_n {\sum\limits_{{l_1},{l_2}} {\phi _{{l_1},{l_2}}^{\left( {k,n} \right)}h_{{l_1},{l_2}}^{\left( {k,n} \right)}} } {q_{{l_1},{l_2}}}{V_{{D_{k,n}}}}{s_k} + \sum\limits_{k \ne k'} {\sum\limits_n {\sum\limits_{{l_1},{l_2}} {\phi _{{l_1},{l_2}}^{\left( {k,n} \right)}h_{{l_1},{l_2}}^{\left( {k,n} \right)}} } {q_{{l_1},{l_2}}}{V_{{D_{k',n}}}}{s_{k'}}}  + {w_k},
\end{equation}
where ${w_k}\sim{\cal{C}\cal{N}}\left( {0,{\sigma ^2}} \right)$ is the additive white Gaussian noise and ${V_{{D_{k,n}}}}$ denotes the $k$-th element in row $n$ of matrix ${{\textbf{V}}_D}$. In $\left(\ref{scalar_signal}\right)$, ${\phi _{{l_1},{l_2}}^{\left( {k,n} \right)}}$ denotes the reflection co-efficient of the RIS element $\left(l_1, l_2\right)$ with respect to the transmitting antenna $n$ and user $k$. In practice, it is a function of the incidence and reflection angels, but here without loss of generality we assume that ${\phi _{{l_1},{l_2}}^{\left( {k,n} \right)}} = {\phi}^{\left(k \right) }$, $\forall n, l_1, l_2$. We ignore the coupling between any two RIS elements here for simplicity, and thus the received signal of each user $k$ comes from the accumulated radiations of all RIS elements, as shown in $\left(\ref{scalar_signal}\right)$. This is a common assumption widely used in the literature on both meta-surfaces~\cite{DOLPN-2017} and traditional antenna arrays~\cite{XAS-2005}.

\subsubsection{Received Signal at the User}
For each user $k$, after it receives the signal $z_k$, it down converts the signal to the baseband and then recovers the final signal. The whole transmission model of $K$ users can be formulated by
\begin{equation}\label{matrix_expression}
{\tilde {\textbf{\emph{z}}}_k} = {{\textbf{F}}}{{\textbf{V}}_D}{\textbf{\emph{s}}} + {\textbf{\emph{w}}},
\end{equation}
where ${\textbf{\emph{w}}} = {\left[ {{w_1}, \cdots ,{w_K}} \right]^T}$ is the noise vector. The transmission matrix ${{\textbf{F}}}$ in $\left( \ref{matrix_expression} \right)$ is defined as
\begin{equation}\label{F_expression}
{\bf{F}} = \sum\limits_{{l_1},{l_2}} {{q_{{l_1},{l_2}}}\left( {{{\bf{H}}_{{l_1},{l_2}}} \circ {{\bf{\Phi }}}} \right)} ,
\end{equation}
where ${{{\bf{H}}_{{l_1},{l_2}}}}$ and ${{\bf{\Phi }}}$ are both $K \times N_t$ matrices consisting of elements $\left\{ {h_{{l_1},{l_2}}^{\left( {k,n} \right)}} \right\}$ and $\left\{ {\phi^{\left(k\right)}} \right\}$, respectively. The notion $\circ$ implies the element-by-element multiplication of two matrices.

\subsection{Sum Rate Maximization Problem Formulation}
To explore how the HBF design influences the sum-rate performance, we evaluate the achievable data rates of all users in the RIS-based system. Based on $\left( \ref{scalar_signal}\right)$ and $\left( \ref{matrix_expression}\right)$, we first rewrite the received signal of user $k$ in matrix form as
\begin{equation}
{z_k} = {{\textbf{F}}_{k}^H}{{\textbf{V}}_{D,k}}{s_k} + \underbrace{\sum\limits_{k' \ne k} {{{\textbf{F}}_{k}^H}} {{\textbf{V}}_{D,k'}}{s_{k'}}}_{\mbox{inter-user interference}} + {w_k},
\end{equation}
where ${\textbf{F}}_{k}$ and ${\textbf{V}}_{D,k}$ denote the $k$-th columns of matrices ${{\textbf{F}}}$ and ${\textbf{V}}_{D}$, respectively. The achievable rate of user $k$ can then be given by
\begin{equation}
R_k = {\log _2}\left( {1 + \frac{{{{\left| {{{\textbf{F}}_{k}^H}{{\textbf{V}}_{D,k}}} \right|}^2}}}{{\sum\limits_{k' \ne k} {{{\left| {{{\textbf{F}}_{k}^H}{{\textbf{V}}_{D,k'}}} \right|}^2}}  + \sigma^2}}} \right).
\end{equation}

We aim to maximize the achievable rates of all users by optimizing the digital beamformer ${\textbf{V}}_D$ and the RIS configuration $\left\{ {{q_{{l_1},{l_2}}}} \right\}$, as formulated below:
\begin{subequations} \label{RIS_problem}
	\begin{align}
	\mathop {\mathrm{maximize}}\limits_{  {{{\textbf{V}}_D}},\left\{ {{q_{{l_1},{l_2}}}} \right\} } \quad &\sum\limits_{1 \le k \le K} {R_k} \\
	{\mathrm{subject\ to}} \quad & {\text{Tr}}\left( {{{\textbf{V}}_D^H}{\textbf{V}}_D} \right) \le P_T,\label{LEO_3}\\
	& q_{l_1, l_2} = \frac{j + e^{j\theta_{{l_1},{l_2}}}}{2}, 0 \le l_1, l_2 \le N_R - 1,\label{RIS_2}\\
	& {\theta _{{l_1},{l_2}}} = \frac{{m_{l_1,l_2}\pi }}{{{2^{b - 1}}}},m_{l_1,l_2} \in\left\lbrace 0,1, \ldots ,{2^b} - 1\right\rbrace,
	%& \sum\limits_{{l_2} = 1}^{{N_R} - 1} {{q_{{l_1},{l_2}}} \cdot {{\left( {{q_{{l_1},{l_2}}}} \right)}^*}} {\rm{  }} = \sum\limits_{{l_2} = 1}^{{N_R} - 1} {{q_{{l_1}',{l_2}}} \cdot {{\left( {{q_{{l_1}',{l_2}}}} \right)}^*}} ,\forall 0 \le {l_1},{l_1}^\prime  \le {N_R} - 1,\label{q_theta}
	\end{align}
\end{subequations}
where $P_T$ is the total transmit power of the BS.

\subsection{Problem Decomposition}
Note that problem $\left(\ref{RIS_problem} \right)$ is a mixed integer non-convex optimization problem which is very challenging due to the large number of discrete variables $\left\{{{q_{{l_1},{l_2}}}} \right\}$ as well as the coupling between propagation and RIS configuration based analog beamforming. Traditional analog beamforming design methods with finite resolution phase shifters~\cite{FW-2016} may not fit well since it is non-trivial to decouple the transmission matrix ${\textbf{F}}$ into the product of a channel matrix and a beamformer matrix in our case. To solve this problem efficiently, we decouple it into two subproblems as shown below.

%\subsection{Upper Bound of the Achievable Sum Rate}
%The achievable rate of all users can be upper bounded by that of an RIS-based system with unlimited surface size and continuous configuration design.

%problems: upper bound (continuous or optimal MIMO); comparison with MIMO or relay

\subsubsection{Digital Beamforming}
Given RIS configuration $\{q_{l_1,l_2}\}$, the digital beamforming subproblem can be written by
\begin{subequations}\label{lab:sub1}
	\begin{align}
	\mathop {\mathrm{maximize}}\limits_{\bm{V}_D}~& \sum\limits_{1 \leq k \leq K} R_k, \\
	\operatorname{subject~to} &~\text{Tr}\left( {{{\textbf{V}}_D^H}{\textbf{V}}_D} \right) \le P_T,
	\end{align}
\end{subequations}
where $\bm{F}$ is fixed.
\subsubsection{RIS Configuration based Analog Beamforming}
Based on constraint $\left(\ref{RIS_2} \right)$, the RIS configuration subproblem with fixed beamformer $\bm{V}_D$ is equivalent to
\begin{subequations}\label{lab:sub2}
	\begin{align}
	\mathop {\mathrm{maximize}}\limits_{\{\theta_{l_1,l_2}\}}~& \sum\limits_{1 \leq k \leq K} R_k, \\
	\operatorname{subject~to} &~{\theta _{{l_1},{l_2}}} = \frac{{m_{l_1,l_2}\pi }}{{{2^{b - 1}}}},m_{l_1,l_2} \in\left\lbrace 0,1, \ldots ,{2^b} - 1\right\rbrace. \label{discrete_constraint}
	%	& \sum\limits_{{l_2} = 0}^{{N_R} - 1} {\left( {1 + \sin {\theta _{{l_1},{l_2}}}} \right)} = \sum\limits_{{l_2} = 0}^{{N_R} - 1} {\left( {1 + \sin {\theta _{{l_1}',{l_2}}}} \right)} ,\forall 0 \le {l_1},{l_1}' \le {N_R} - 1.
	\end{align}
\end{subequations}

In the next section, we will design two algorithms to solve these subproblems, respectively.

\section{Sum Rate Maximization Algorithm Design}
In this section, we will develop a sum rate maximization~(SRM) algorithm to obtain a suboptimal solution of problem (\ref{RIS_problem}) in Section III. Specially, we iteratively solve subproblem (\ref{lab:sub1}) given RIS configuration $\{q_{l_1,l_2}\}$, and solve subproblem (\ref{lab:sub2}) given beamformer $\bm{V}_D$. Finally, we will summarize the overall algorithm and provide convergence and complexity analysis.

\subsection{Digital Beamforming Algorithm}

Subproblem (\ref{lab:sub1}) is a well-known digital beamforming problem. According to the results in~\cite{FDBETOF-2013}, the ZF digital beamformer can obtain a near optimal solution. Therefore, we consider ZF beamforming together with power allocation as the beamformer at the BS to alleviate the interference among users. Based on the results in \cite{CBA-2005}, the beamformer can be given by
\begin{equation}\label{precoder}
\bm{V}_D = \bm{F}^H(\bm{F}\bm{F}^H)^{-1}\bm{P}^{\frac{1}{2}} = \tilde{\bm{V}}_D \bm{P}^{\frac{1}{2}},
\end{equation}
where $\tilde{\bm{V}}_D = \bm{F}^H(\bm{F}\bm{F}^H)^{-1}$ and $\bm{P}$ is a diagonal matrix whose $k$-th diagonal element is the received power at the $k$-th user, i.e., $p_k$.

In the ZF beamforming, we have the following constraints:
\begin{equation}
\begin{array}{ll}
&|\bm{F}_k^{H}(\bm{V}_D)_k| = \sqrt{p_k}, \\
&|\bm{F}_k^H (\bm{V}_D)_{k'}| = 0, \forall k' \ne k.
\end{array}
\end{equation}
With these constraints, subproblem (\ref{lab:sub1}) can be reduced to the following power allocation problem:
\begin{subequations}\label{power_allocation}
	\begin{align}
	\max\limits_{\{p_k \geq 0\}}~& \sum\limits_{1 \leq k \leq K} \log_2\left(1 + \frac{p_k}{\sigma^2}\right), \\
	\operatorname{subject~to} &~\text{Tr}\left( {{{\bm{P}}^{\frac{1}{2}}}\tilde {\bm{V}}_D^H{{\tilde {\bm{V}}}_D}{{\bm{P}}^{\frac{1}{2}}}} \right) \le {P_T} \label{cons1}.
	\end{align}
\end{subequations}
The optimal solution of this problem can be obtained by water-filling~\cite{DP-2005} as
\begin{equation}\label{solution}
p_k = \frac{1}{\nu_k}\max\left\lbrace \frac{1}{\mu}-\nu_k\sigma^2,0\right\rbrace ,
\end{equation}
where $\nu_k$ is the $k$-th diagonal element of $\tilde{\bm{V}}_D^H\tilde{\bm{V}}_D$ and $\mu$ is a normalized factor which is selected such that $\sum \limits_{1 \leq k \leq K}\max\{\frac{1}{\mu}-\nu_k\sigma^2,0\} = P_T$. The algorithm can be summarized in Algorithm \ref{BF_Algorithm}.

\begin{algorithm}[!t]
	%\small
	\caption{Digital Beamforming Algorithm}\label{BF_Algorithm}
	\begin{algorithmic}[1]
		\State Solve power allocation problem (\ref{power_allocation});
		\State Obtain the optimal power allocation result (\ref{solution});
		\State Derive the beamformer matrix from the optimal power allocation based on (\ref{precoder});
	\end{algorithmic}
\end{algorithm}

\subsection{RIS Configuration based Analog Beamforming Algorithm}
Since we iterate between the digital beamforming and RIS configuration based analog beamforming, the latter can be optimized assuming ZF precoding as shown in (\ref{precoder}). Since the data rate with ZF precoding in (\ref{power_allocation}) only depends on the RIS configuration through the power constraint (\ref{cons1}), the RIS configuration based analog beamforming problem can be reformulated as a power minimization problem:
\begin{subequations}
	\begin{align}
	\min\limits_{\theta_{l_1,l_2}} & f(\bm{F}),\\
	\operatorname{subject~to} &\theta_{l_1,l_2} = \frac{m_{l_1,l_2}\pi}{2^{b-1}},m_{l_1,l_2} \in \{0,1,\ldots,2^b -1\},
	\end{align}
\end{subequations}
where
\begin{equation}
\begin{array}{ll}
f(\bm{F}) & = \text{Tr}(\tilde{\bm{V}}_D \bm{P} \tilde{\bm{V}}_D^H) = \text{Tr}(\bm{P}^{\frac{1}{2}} \tilde{\bm{V}}_D^H \tilde{\bm{V}}_D \bm{P}^{\frac{1}{2}})\\
& = \text{Tr}((\tilde{\bm{F}}\tilde{\bm{F}}^H)^{-1}).
\end{array}
\end{equation}
Here, $\tilde{\bm{F}} = \bm{P}^{-\frac{1}{2}}\bm{F}$.

Since $\tilde{\bm{F}}\tilde{\bm{F}}^H$ is a symmetric, positive semi-definite matrix, we can transform this problem into a semi-definite programming (SDP) problem. Let $\text{Tr}((\tilde{\bm{F}}\tilde{\bm{F}}^H)^{-1}) = \text{Tr}(\frac{w}{K} \bm{I}_K)$. According to Schur complement \cite{J-2010}, the problem can be rewritten by
\begin{subequations}\label{lab:sub3}
	\begin{align}
	\min\limits_{\theta_{l_1,l_2},w} & w, \\
	\operatorname{subject~to} &~ \bm{Z} = \left[\begin{matrix}\frac{w}{K} \bm{I}_K & \bm{I}_K\\ \bm{I}_K & \tilde{\bm{F}}\tilde{\bm{F}}^H \end{matrix}\right] \succeq 0, \label{matrix_semi}\\
	%&\sum\limits_{l_2 = 0}^{N_r-1}(\sin \theta_{l_1,l_2} - \sin \theta_{{l_1}',l_2}) = 0, \label{sin}\\
	&\theta_{l_1,l_2} = \frac{m_{l_1,l_2}\pi}{2^{b-1}},m_{l_1,l_2} \in \{0,1,\ldots,2^b -1\},
	\end{align}
\end{subequations}
where $\bm{X} \succeq 0$ means that matrix $\bm{X}$ is a symmetric and positive semi-definite matrix.

\emph{Remark on the tractability of the formulated problem}: This problem is a mix-integer SDP, which is generally NP-hard. Moreover, any two discrete variables $\left\lbrace \theta_{l_1,l_2}\right\rbrace$ are coupled with each other via constraint $\left( \ref{matrix_semi} \right)$, which makes the problem even more complicated. One commonly used solution is to first relax the discrete variables into continuous ones and then round the obtained solution to satisfy the discrete constraints. However, for the RIS-based systems, the typical value of the number of quantization bits is usually very small (e.g., 2 or 3) such that the round-off methods will lead to inevitable performance degrade.

To avoid the above issue, we consider to solve the SDP discretely. In the following, we first present the following Proposition~\ref{pro_3} to transform the nonlinear functions in~$\left( \ref{lab:sub3}\right)$ with respect to $\theta_{l_1,l_2}$ into linear ones, as proved in Appendix~\ref{proof_3}. We then use the outer approximation method \cite{TMS-2018} to solve this problem.

\begin{proposition}\label{pro_3}
	Let
	\begin{align}
	&\bm{a} = \left[-\frac{(2^{b} - 1)\pi}{2^{b-1}},\ldots,-\frac{m\pi}{2^{b-1}},\ldots, \frac{m\pi}{2^{B-1}}, \ldots, \frac{(2^{b} - 1)\pi}{2^{b-1}} \right],\nonumber\\
	&\bm{c} = \left[\cos\left(-\frac{(2^{b} - 1)\pi}{2^{b-1}}\right),\ldots,  \cos\left(-\frac{m\pi}{2^{b-1}}\right), \ldots, \cos\left(\frac{m\pi}{2^{b-1}}\right), \ldots, \cos\left(\frac{(2^{b} - 1)\pi}{2^{b-1}}\right) \right],\nonumber\\
	&\bm{s} = \left[\sin\left(-\frac{(2^{b} - 1)\pi}{2^{b-1}}\right),\ldots, \sin\left(-\frac{m\pi}{2^{b-1}}\right), \ldots, \sin\left(\frac{m\pi}{2^{b-1}}\right), \ldots, \sin\left(\frac{(2^{b} - 1)\pi}{2^{b-1}}\right) \right].\nonumber
	\end{align}
	 We introduce a binary vector $\bm{x}^{l_1,l_2}$, where $x_i^{l_1,l_2}$ indicates whether $\theta_{l_1,l_2} = a_{i}$, and a binary vector $\bm{y}^{l_1,l_2,{l_1}',{l_2}'}$ for phase difference $\Delta \theta_{l_1,l_2,{l_1}',{l_2}'} = \theta_{l_1,l_2} - \theta_{{l_1}',{l_2}'}$. Therefore, problem (\ref{lab:sub3}) can be rewritten by
	\begin{subequations}\label{lab:sub4}
		\begin{align}
		\min\limits_{\bm{x}^{l_1,l_2},\bm{y}^{l_1,l_2,{l_1}',{l_2}'},w} & w, \\
		\operatorname{subject~to} &~ \bm{Z} = \left[\begin{matrix}\frac{w}{K} \bm{I}_K & \bm{I}_K\\ \bm{I}_K & \tilde{\bm{F}}\tilde{\bm{F}}^H \end{matrix}\right] \succeq 0, \\
		&\|\bm{x}^{l_1,l_2}\|_1 = 1, \bm{e}^T\bm{x}^{l_1,l_2} = 0,\\
		&\bm{a}^{T}(\bm{x}^{l_1,l_2} - \bm{x}^{{l_1}',{l_2}'}) = \bm{a}^{T}\bm{y}^{l_1,l_2,{l_1}',{l_2}'}.
		\end{align}
	\end{subequations}
	Here, $\bm{e}$ is a constant vector whose first $2^B - 1$ elements are 1 and others are 0.
\end{proposition}

%$\bm{a} = \left[-\frac{(2^{b} - 1)\pi}{2^{b-1}},\ldots,-\frac{m\pi}{2^{b-1}},\ldots, 0, \ldots, \frac{m\pi}{2^{B-1}},\right.\\\left. \ldots, \frac{(2^{b} - 1)\pi}{2^{b-1}} \right]$, $\bm{c} = \left[\cos\left(-\frac{(2^{b} - 1)\pi}{2^{b-1}}\right),\ldots,  \cos\left(-\frac{m\pi}{2^{b-1}}\right), \ldots, \cos(0), \ldots, \cos\left(\frac{m\pi}{2^{b-1}}\right), \ldots, \cos\left(\frac{(2^{b} - 1)\pi}{2^{b-1}}\right) \right]$, and $\bm{s} = \left[\sin\left(-\frac{(2^{b} - 1)\pi}{2^{b-1}}\right),\ldots, \sin\left(-\frac{m\pi}{2^{b-1}}\right), \ldots, \sin(0), \ldots, \sin\left(\frac{m\pi}{2^{b-1}}\right), \ldots, \sin\left(\frac{(2^{b} - 1)\pi}{2^{b-1}}\right) \right]$.

Problem (\ref{lab:sub4}) is a mix-integer SDP with linear constraints, which can be solved by the outer approximation method. The basic idea of the outer approximation method is to enforce the SDP constraint via linear cuts and transform the original problem into a mix-integer linear programming one, which can be solved by the branch-and-bound algorithm \cite{SHBL-2019}. In the following, we will elaborate on how to enforce the SDP constraints via linear cuts.

Assume that a solution is $\bar{w},\bar{\bm{x}}^{l_1,l_2},\bar{\bm{y}}^{l_1,l_2,{l_1}',{l_2}'}$. In most mix-integer programming problems, it is very common to use the gradient cuts to approach the feasible set. However, the function of smallest eigenvalues is not always differentiable. Therefore, we use the characterization instead \cite{TMS-2018}. Note that $\bm{Z} \succeq 0$ is equivalent to $\bm{u}^{T}\bm{Z}\bm{u} \geq 0$ for arbitrary $\bm{u}$. If $\bm{Z}$ with $\bar{w},\bar{\bm{x}}^{l_1,l_2},\bar{\bm{y}}^{l_1,l_2,{l_1}',{l_2}'}$ is not positive semi-definite, we compute eigenvector $\bm{u}$ associated with the smallest eigenvalue. Then
\begin{equation}\label{cut}
\bm{u}^{T}\bm{Z}\bm{u} \geq 0
\end{equation}
is a valid cut that cuts off $\bar{w},\bar{\bm{x}}^{l_1,l_2},\bar{\bm{y}}^{l_1,l_2,{l_1}',{l_2}'}$. The RIS configuration based analog beamforming algorithm can be summarized in Algorithm \ref{RIS_Algorithm}.

\begin{algorithm}[!t]
	%\small
	\caption{RIS Configuration based Analog Beamforming Algorithm}\label{RIS_Algorithm}
	\begin{algorithmic}[1]
		
		\State Remove the semi-definite constraint of problem (\ref{lab:sub4}) and solve an initial solution;
		\Repeat
		\State Use the branch-and-bound method to solve problem (\ref{lab:sub4}) and obtain the optimal solution $\bm{x}^{l_1,l_2}$ for each RIS element;
		\State Check the feasibility;
		\State If the obtained solution is not feasible, add a cut according to (\ref{cut});
		\Until{The obtained solution is feasible;}
		\State Derive the phase shits according to the obtained solution.
	\end{algorithmic}
\end{algorithm}

\subsection{Overall Algorithm Description}
Based on the results presented in the previous two subsections, we propose an overall iterative algorithm, i.e., the SRM algorithm, for solving the original problem in an iterative manner. Specially, the beamformer $\bm{V}_D$ is solved by Algorithm \ref{BF_Algorithm} while keeping the RIS configuration fixed. After obtaining the results, we will optimize the RIS configuration $\theta^{l_1,l_2}$ by Algorithm \ref{RIS_Algorithm}. Those obtained results are set as the initial solution for subsequent iterations. Define $R$ as the value of the objective function. The two subproblems will be solved alternatively until in iteration $t$ the value difference of the objective functions between two adjacent iterations is less than a predefined threshold $\pi$, i.e., $R^{(t +1)} - R^{(t)} \leq \pi$.

\subsection{Convergence and Complexity Analysis}
We now analyze the convergence and complexity of our proposed SRM algorithm.

\subsubsection{Convergence}
First, according to Algorithm 1, in the digital beaforming subproblem, we can obtain a better result given RIS configuration $\bm{\theta}^{(t)}$ in the $\left( t+1\right)$-th iteration. Therefore, we have
\begin{equation}
R(\bm{V}_D^{(t + 1)}, \bm{\theta}^{(t)}) \geq R(\bm{V}_D^{(t)}, \bm{\theta}^{(t)}).
\end{equation}
Second, given the beamforming result $\bm{V}_D^{(t + 1)}$, we maximize sum rate of all users, and thus, the following inequality holds:
\begin{equation}
R(\bm{V}_D^{(t + 1)}, \bm{Q}^{(t + 1)}) \geq R(\bm{V}_D^{(t + 1)}, \bm{\theta}^{(t)}).
\end{equation}
Based the above inequalities, we can obtain
\begin{equation}
R(\bm{V}_D^{(t + 1)}, \bm{\theta}^{(t + 1)}) \geq R(\bm{V}_D^{(t)}, \bm{Q}^{(t)}).
\end{equation}
which implies that the objective value of the original problem is non-decreasing after each iteration of the SRM algorithm. Since the objective value is upper bounded, the proposed SRM algorithm is guaranteed to converge.

\subsubsection{Complexity}
We consider the complexity of the proposed algorithms for two subproblems separately.
\begin{itemize}
	\item In the digital beamforming subproblem, we need to optimize the received power for each user according to (\ref{solution}). Therefore, its computational complexity is $O(K)$.
	\item In the RIS configuration based analog beamforming subproblem, we solve a series of linear programs by the branch-and-bound method. Since only one element in $\bm{x}^{l_1,l_2}$ can be 1, $\bm{x}^{l_1,l_2}$ can have at most $2^b$ possible solutions. Thus, the scale of the computational complexity of each linear program is $O(2^{bN_T^2})$.
\end{itemize}

\section{Performance Analysis of RIS-based Multi-user Communications}
In this section, we compare the RIS-based HBF scheme with the traditional ones in terms of the minimum number of required RF chains. A special case, i.e., the pure LoS transmission, is also considered to explore theoretically how the size of RIS and its placement influence the achievable rates.

\subsection{Comparison with Traditional Hybrid Beamforming}
We adopt the fully digital beamforming scheme as a benchmark to compare the traditional and RIS-based HBF schemes. In the traditional HBF, it has already been proved that when the number of RF chains is not smaller than twice the number of target data streams, any fully digital beamforming matrix can be realized~\cite{FW-2016}. However, in the RIS-based HBF, the inherent analog beamforming (i.e., the RIS configuration) is closely coupled with propagation, which offers more freedom for shaping the propagation environment than the traditional scheme. To capture this characteristic, we explore a new condition for the RIS-based system to achieve fully digital beamforming.

%the condition to achieve fully digital beamforming can be relaxed owing to the coupling between RIS reconfiguration and propagation.
%Now we show that a similar statement holds in our considered RIS-based system though the analog beamforming is not performed at the BS but by the RIS inherently instead.
We start by describing the fully digital beamforming scheme in an RIS-based system. Consider an ideal case where each RIS element directly connects with an RF chain and ADC as if it is part of the BS\footnote{In practice, the RIS does not connect to the RF chain directly unless it is installed at the BS, which is not the truth in our case where RIS actually only reflects signals. Therefore, we only consider such an ideal scheme as a benchmark to evaluate the effectiveness of RIS-based HBF}. The fully digital beamformer can then be denoted by ${V_{FD}} \in {^{N_R^2 \times K}}$, based on which we present the following proposition, which will be proved in Appendix \ref{proof_prop1}.

\begin{proposition} \label{prop1}
	For the RIS-based HBF scheme with $N_R^2 \ge K{N_t}$, to achieve any fully digital beamforming scheme, the number of transmit antennas at the BS should not be smaller than the number of single-antenna users, i.e., $N_t \ge K$.
\end{proposition}

This implies two conditions for the proposed RIS-based scheme to achieve the fully digial beamforming scheme. First, the size of RIS should be no smaller than the product of the number of users and the size of the antenna array at the BS. Second, the number of transmit antennas at the BS should be no smaller than the number of single-antenna users.

\emph{Remark on dedicated hardware reduction for analog beamforming}: In the traditional HBF, to offload part of the digital baseband processing to the analog domain, a number of RF chains are required at the BS to feed the analog beamformer, equipped with necessary hardwares such as mixers, filters, and phase shifters~\cite{AJNR-2014}. In contrast, as shown in Proposition~\ref{prop1}, the number of minimum RF chains required by the RIS-based HBF to achieve the fully digial beamforming has already been reduced by half compared to the traditional scheme. Moreover, the phase shifters can be saved since the RIS inherently realizes analog beamforming owning to its flexible physical structure~\cite{MAMMP-2019}.

%Such a characteristic of RIS contributes to the unique form of transmission matrix $\textbf{F}$ as shown in $\left( \ref{F_expression}\right)$ combining both phase-shift reconfiguration and reflection-dominated transmission, which is different from those traditional analog beamformers.

\subsection{Special Case: Pure Line-of-Sight Transmissions}
We consider the data rate obtained by the pure LoS case as a lower bound of the achievable rate and analyze optimal RIS placement to provide orthogonal communication links. The LoS case also reveals insights on how the achievable data rate is influenced by RIS design and placement.

Since a multitude of RIS elements are placed in a sub-wavelength order, spacial correlation between these elements are inevitable. In this case, the channel matrix $\textbf{F}$ approaches a low-rank matrix, leading to a degraded performance in terms of the achievable data rate. Traditionally, we can utilize the multi-path effect to decorrelate different channel links between the transceiver antennas. However, it is also important to understand how the system works in the pure LoS case, especially when it comes to the RIS-based systems where the reflection-based one-hop link between the BS and users acts as the dominated ``LoS" component and is usually much stronger than other multi-paths as well as the degraded direct links.

In the pure LoS case, we aim to achieve a high-rank LoS channel matrix by designing the size of RIS and the antenna array at the BS. Different from those existing works on antenna array design for LoS MIMO systems, the discrete phase shifts depicted by $\left\lbrace q_{l_1, l_2} \right\rbrace$ needs to be considered in the RIS-based propagation. We thus present the results in Proposition \ref{prop2} below, which is proved in Appendix \ref{proof_prop2}.

\begin{proposition} \label{prop2}
	In the RIS-based system, to make different links between the BS and user $k$ via the RIS orthogonal to each other, the RIS design should satisfy the following conditions:
	\begin{subequations} \label{RIS_design}
		\begin{align}
		&d_R^{\left( 1 \right)}{d_B}  = \frac{{\lambda D_{0,0}^{\left( 0 \right)}}}{{{N_R}\cos {\theta _R}\cos {\theta _B}}},\label{Rayleigh_rule}\\
		&\sum\limits_{{l_2} = 0}^{{N_R} - 1} {\left( {1 + \sin {\theta _{{l_1},{l_2}}}} \right)}   = \sum\limits_{{l_2} = 0}^{{N_R} - 1} {\left( {1 + \sin {\theta _{{l' _1},{l_2}}}} \right)} ,\forall 0 \le {l_1},{l' _1} \le {N_R} - 1. \label{theta_equal}
		\end{align}
	\end{subequations}
\end{proposition}

This proposition shows that the achievable data rate is highly related to the size of the RIS and its placement. For convenience, when all other parameters are fixed, we refer to the \emph{threshold} of the RIS size as
\begin{equation}\label{threshold_size}
N_R^{th} = \frac{{\lambda D_{0,0}^{\left( 0 \right)}}}{{d_R^{\left( 1 \right)}{d_B}\cos {\theta _R}\cos {\theta _B}}}.
\end{equation}

For the pure LoS case, the sum rate maximization problem can still be formulated as $\left( \ref{RIS_problem} \right)$ with one extra constraint $\left( \ref{theta_equal} \right)$. Our proposed SRM algorithm can be utilized to solve this problem after we convert the extra constraint $\left( \ref{theta_equal} \right)$ into a linear one shown below by following the transformations in Proposition 1,
\begin{equation}
\sum\limits_{l_2 = 0}^{N_r-1}(\bm{x}^{l_1,l_2} - \bm{x}^{{l_1}',l_2})\bm{s}^{T} = 0.
\end{equation}

%\sum\limits_{l_2 = 0}^{N_r-1}(\sin \theta_{l_1,l_2} - \sin \theta_{{l_1}',l_2}) = 0,\\

%&\sum\limits_{l_2 = 0}^{N_r-1}(\bm{x}^{l_1,l_2} - \bm{x}^{{l_1}',l_2})\bm{s}^{T} = 0,\\

\section{Simulation Results}
In this section, we evaluate the performance of our proposed algorithm for RIS-based HBF in terms of the sum rate. We show how the system performance is influenced by the SNR, number of users, the size of RIS, and the number of quantization bits for discrete phase shifts. For comparison, the following algorithms are performed as well.
\begin{itemize}
	\item \emph{Simulated annealing}: We utilize the simulated annealing method~\cite{BLY-2016} to approach the global optimal solution of the sum rate maximization problem with discrete phase shifts. The maximum number of iterations is set as ${10^7}$.
	\item \emph{Pure LoS case}: We consider the pure LoS case as a lower bound to evaluate the performance of the HBF scheme. The proposed iterative SRM algorithm can still be utilized to solve the corresponding optimization problem~.
	\item \emph{Random phase shift}: Algorithm 1 is first performed, followed by a random algorithm to solve the RIS-based analog beamforming subproblem $\left( \ref{lab:sub2} \right)$.
	\item \emph{RIS-based HBF with continuous phase shifts}: This scheme only serves as a benchmark when we investigate how the discreteness level (i.e., the number of quantization bits, $b$) influences the sum rate. The discrete constraint in the original subproblem $\left( \ref{lab:sub2} \right)$ is relaxed to $\theta_{{l}_1,{l}_2} \in \left[ 0, 2\pi\right]$. The HBF solution is obtained by iteratively performing Algorithm~\ref{BF_Algorithm} and the gradient descent method.
\end{itemize}

\begin{figure}[!t]
	\centering
	\includegraphics[width=5in]{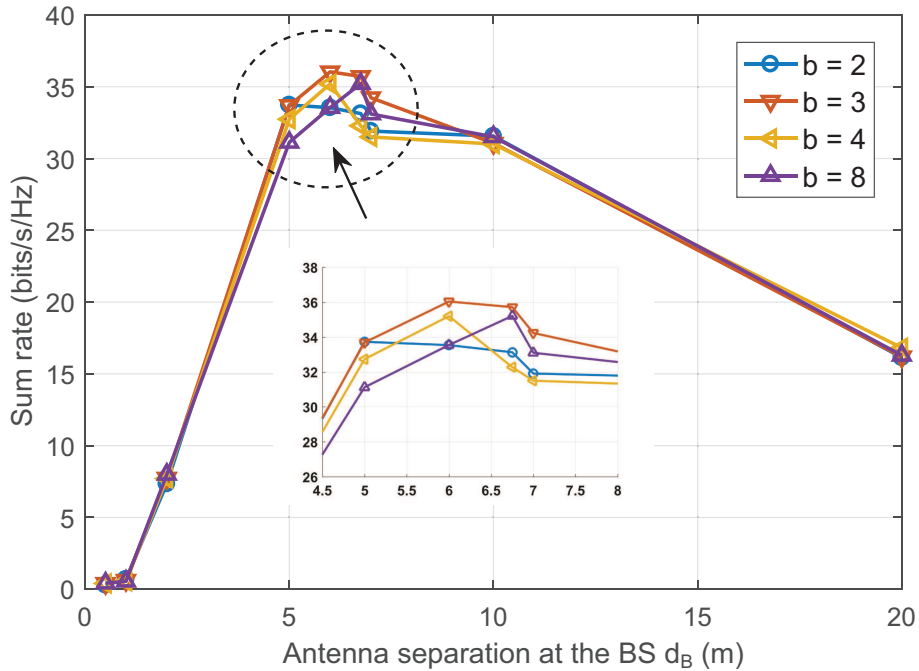}
	%\vspace{-0.5cm}
	\caption{Sum rate v.s. antenna separation at the BS (SNR = 2 dB, $N_t = K = 5$, $N_R = 6$)} \label{ULA}
	%\vspace{-0.9cm}
\end{figure}

In our simulation, we set the distance between the BS and the RIS, $D_{0,0}^{\left( 0 \right)}$, as 20m, and users are randomly deployed within a half circle of radius 60m centering at the RIS. The antenna array at the BS and the RIS are placed at angles of ${\rm{1}}{{\rm{5}}^ \circ }$ and ${\rm{3}}{{\rm{0}}^ \circ }$ to the x axis, respectively. The transmit power of the BS $P_T$ is 20 W, the carrier frequency is 5.9 GHz, the antenna separation at the BS $d_B$ is 1 m, the RIS element separation $d_R$ is 0.03 m, and the Rician fading parameter $\kappa$ is 4 \cite{HMK-2014}. We set the size of the RIS $N_R^2$ ranging between ${5^2} \sim {65^2}$, the number of antennas at the BS $N_t$ and the number of  users $K$ between 5 $\sim$ 15, the discreteness level of RIS $b$ between 1 $\sim$ 5, and the SNR (defined as $P_T/{\sigma^2})$ between -2 dB $\sim$ 10 dB. Specifically, for the LoS case, given the above parameters the designing rule in $\left( \ref{Rayleigh_rule} \right)$ is only satisfied when the size of RIS is set as 40 or the antenna separation at the BS is set as 6.75 m.

%\begin{figure}
%	\centering
%	\subfigure[]{
%		\label{upa_separation} %% label for first subfigure
%		\includegraphics[width=3.2in]{upa_separation.eps}}
%	\hspace{-0.2in}
%	\subfigure[]{
%		\label{ula_separation} %% label for second subfigure
%		\includegraphics[width=3.2in]{ula_separation.eps}}
%	\caption{a) Sum rate v.s. element separation at the RIS (SNR = 2 dB, $K = N_t = 4$, $N_R = 6$); b) Sum rate v.s. antenna separation at the BS (SNR = 2 dB, $N_t = K = 4$, $N_R = 6$)} \label{separation} %% label for entire figure
%\end{figure}

We present Fig.~\ref{ULA} to verify Proposition~\ref{prop2} in the pure LoS case. The figure shows the sum rate of all users versus the antenna separation at the BS, $d_B$, with different numbers of quantization bits in the pure LoS case. Given the parameters set above, according to Proposition~\ref{prop2},  the optimal value of $d_B$ should be 6.75 m so as to orthogonalize the channel links between each antenna of the BS and a user $k$ via any RIS element. We observe that the optimal sum rate can be achieved when $d_B$ is around $5 \sim 6.75$ m. The numerically optimal value of $d_B$ approaches the theoretical result as the number of quantization bits, $b$, grows, implying that such fluctuation around the optimal value comes from the discrete phase shifts of RIS. When $b$ is large enough, i.e., $b = 8$ in Fig.~\ref{ULA}, the optimal $d_B$ equals 6.75 m, which justifies Proposition~\ref{prop2}.

%Similarly, in Fig.~\ref{ula_separation}, we observe that the numerically optimal value of $d_B$ equals the theoretical value 6.75 m when $b = 8$. This also indicates that given all other parameters, there exists an optimal RIS element separation to maximize the achievable rate in the pure LoS case, which will serve as a reference for RIS design in more general LoS-dominated cases.

\begin{figure}[!t]
	\centering
	\includegraphics[width=5in]{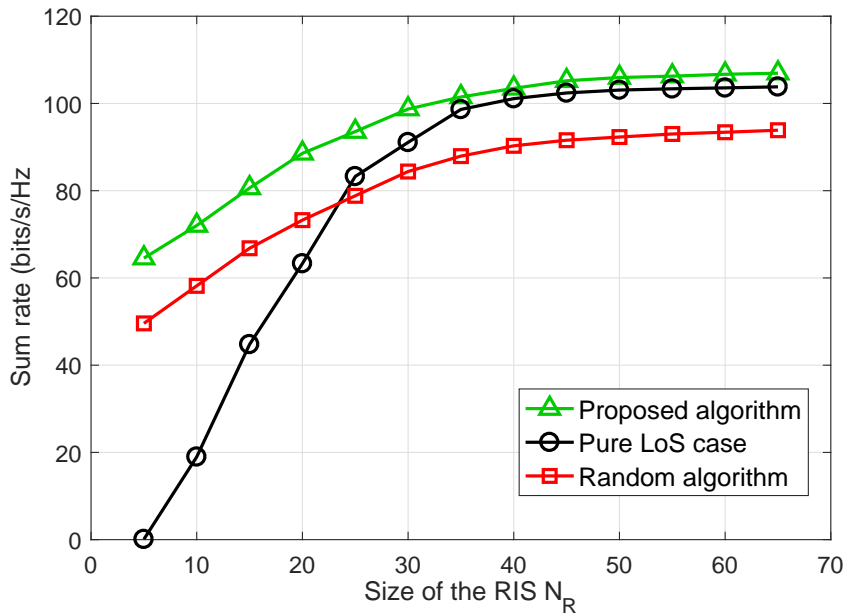}
	%\vspace{-0.5cm}
	\caption{Sum rate v.s. size of RIS $N_R$ (SNR = 2dB, $K = N_t = 5$, $b = 2$)} \label{size_rate}
	%\vspace{-0.9cm}
\end{figure}

Fig.~\ref{size_rate} shows the sum rate of all users versus the size of RIS\footnote{For convenience, here we adopt $N_R$ to represent the size of RIS to better display the curves.} with $b = 2$, $N_t = K = 5$. We observe that the sum rate grows rapidly with a small size of RIS and gradually flattens as the size of RIS continues to increase\footnote{We do not show the simulated annealing algorithm in this figure due to its high complexity with a large size of RIS.}. The inflection point of each curve shows up around $N_R = 40$, which verifies the threshold $\left( \ref{threshold_size} \right)$ given by Proposition~\ref{prop2} very well. When the RIS size $N_R$ exceeds 40, though Proposition~\ref{prop2} does not hold, the sum rate does not drop since the RIS can always turn off those extra RIS elements to maintain the sum-rate performance. This figure also implies that though Proposition~\ref{prop2} is obtained in the pure LoS case, it also sheds insight into the RIS placement and array design in a more general case with small-scale fading.

\begin{figure}[!t]
	\centering
	\includegraphics[width=5in]{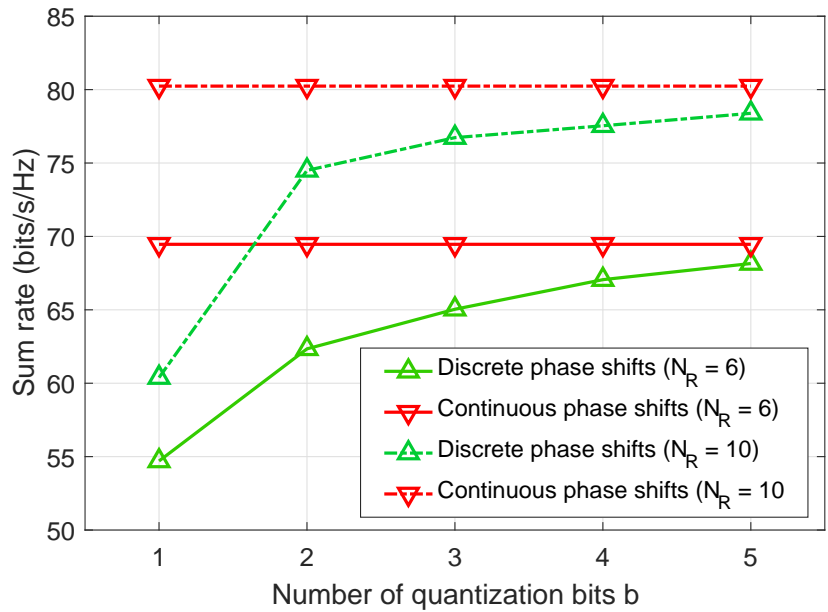}
	%\vspace{-0.5cm}
	\caption{Sum rate v.s. number of quantization bits $b$ (SNR = 2dB, $K = N_t = 5$, $N_R = 6$)} \label{bit_rate}
	%\vspace{-0.9cm}
\end{figure}

Moreover, Fig.~\ref{size_rate} also shows that the performance of RIS-based beamforming with small-scale fading is much better than that of the pure LoS case when the size of RIS is small. Such a gain comes from the reduced correlation between different channel links owning to multi-path effects. As the size of RIS grows, Proposition~\ref{prop2} is satisfied such that the channel links in the pure LoS case are orthogonalized, making the gap between these two cases smaller.

Fig.~\ref{bit_rate} depicts the sum rate of all users versus the number of quantization bits $b$ for discrete phase shifts in RIS configuration with SNR = 2 dB, $N_t = K = 5$, and $N_R = 6$. As the number of quantization bits increases, the sum rate obtained by our proposed algorithm with discrete phase shifts approaches that in the continuous case. When the size of RIS grows, the gap between the discrete and continuous cases shrinks since a larger RIS usually provides more freedom of generating directional beams. Note that the implementation difficulty increases dramatically in practice with the number of quantization bits. A trade-off can then be achieved between the sum rate and number of quantization bits.

%\begin{figure}[!t]
%	\centering
%	\includegraphics[width=4.8in]{SNR_rate.eps}
%	%\vspace{-0.5cm}
%	\caption{Sum rate v.s. SNR ($K = N_t = 5$, $b = 2$, $N_R = 6$)} \label{snr_rate}
%	%\vspace{-0.9cm}
%\end{figure}

\begin{figure}
	\centering
	\subfigure[]{
		\label{snr_rate} %% label for first subfigure
		\includegraphics[width=3.2in]{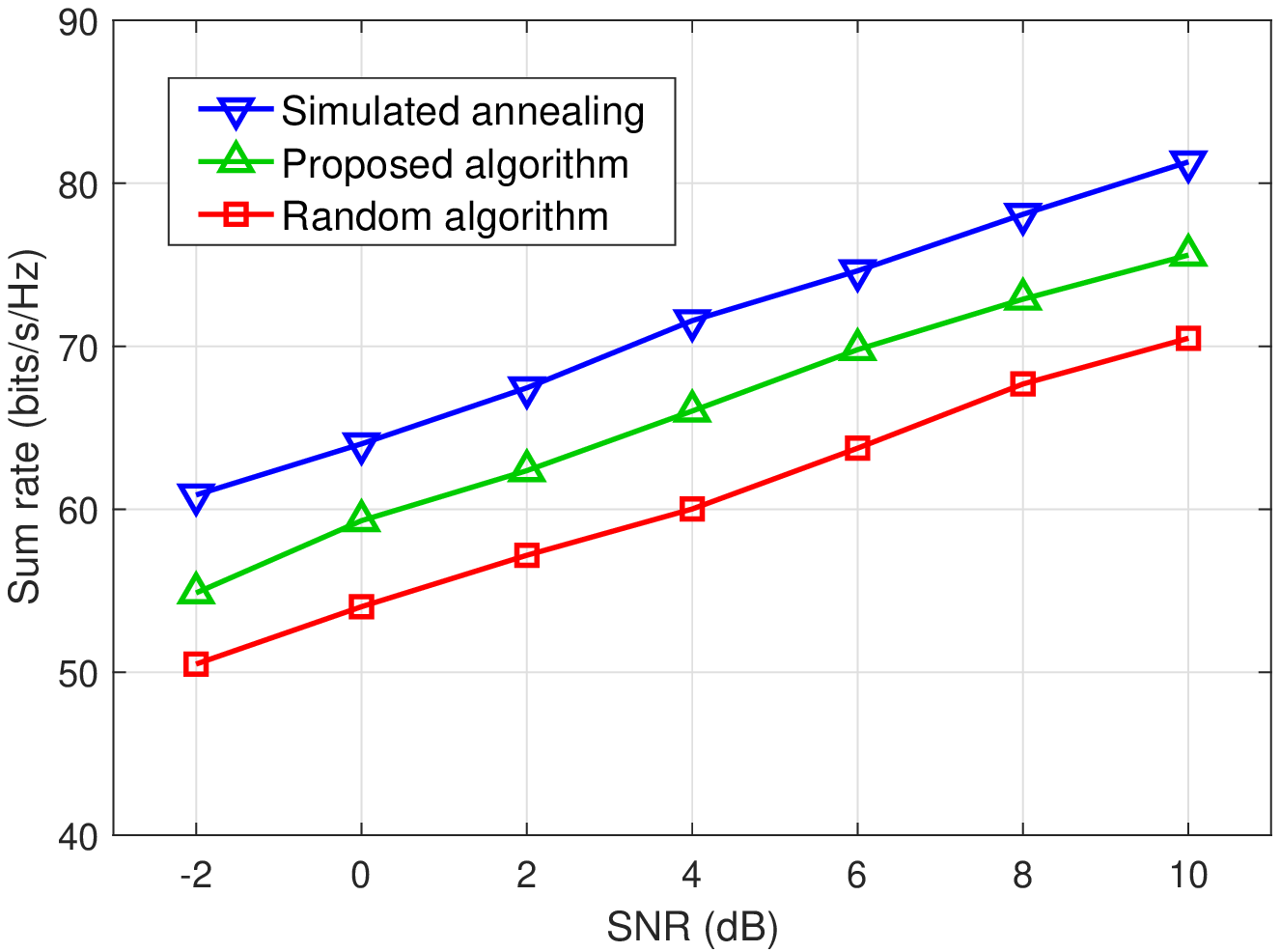}}
	\hspace{-0.2in}
	\subfigure[]{
		\label{user_num_rate} %% label for second subfigure
		\includegraphics[width=3.2in]{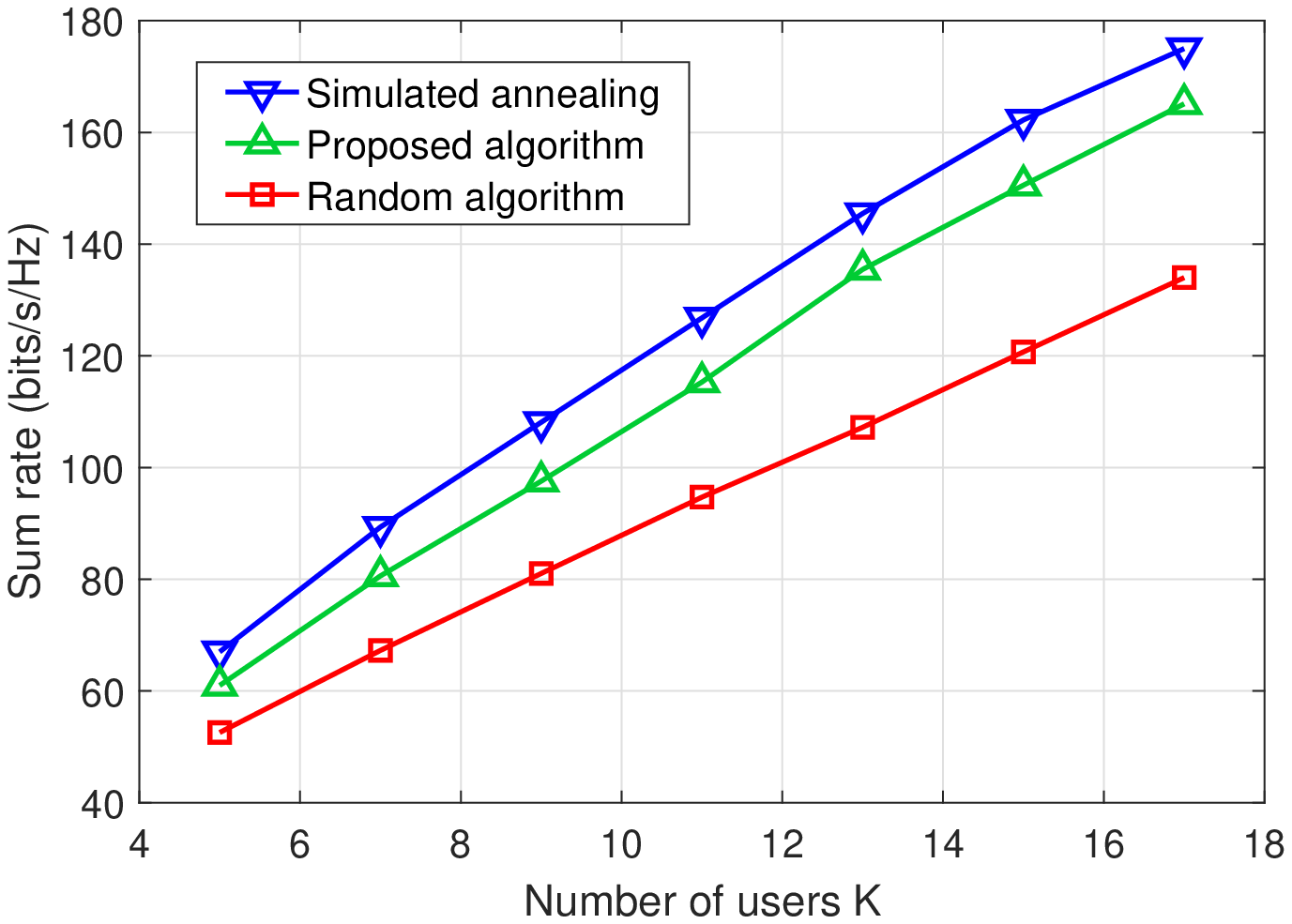}}
	\caption{a) Sum rate v.s. SNR ($K = N_t = 5$, $b = 2$, $N_R = 6$); b) Sum rate v.s. number of users (SNR = 2 dB, $N_t = K$, $b = 2$, $N_R = 6$)} \label{rate_increase} %% label for entire figure
\end{figure}

%\begin{figure}[!t]
%	\centering
%	\includegraphics[width=4.6in]{SNR_rate.eps}
%	%\vspace{-0.5cm}
%	\caption{Sum rate v.s. SNR ($K = N_t = 5$, $b = 2$, $N_R = 6$)} \label{snr_rate}
%	%\vspace{-0.9cm}
%\end{figure}
%
%\begin{figure}[!t]
%	\centering
%	\includegraphics[width=5in]{user_num_rate.eps}
%	%\vspace{-0.5cm}
%	\caption{Sum rate v.s. number of users (SNR = 2 dB, $N_t = K$, $b = 2$, $N_R = 6$)} \label{user_num_rate}
%	%\vspace{-0.9cm}
%\end{figure}

Fig.~\ref{snr_rate} and Fig.~\ref{user_num_rate} show the sum rate of all users versus SNR and the number of users, respectively, obtained by different algorithms with a RIS of size $6 \times 6$ (i.e., $N_R = 6$), $b = 2$ quantization bits for phase shifts, equal number of transmit antennas at the BS and the downlink users. In Fig.~\ref{snr_rate}, the sum rate increases with SNR since more power resources are allocated by the BS. In Fig.~\ref{user_num_rate}, the sum rate grows with the number of users since a higher diversity gain is achieved. From both figures, we observe that the performance of our proposed algorithm is close to that of the simulated annealing method and much better than the random algorithm. This indicates the efficiency of our proposed algorithm to solve the RIS-based HBF problem.

%\begin{figure}[!t]
%	\centering
%	\includegraphics[width=4.6in]{user_num_rate.eps}
%	%\vspace{-0.5cm}
%	\caption{Sum rate v.s. number of users (SNR = 2 dB, $N_t = K$, $b = 2$, $N_R = 6$)} \label{user_num_rate}
%	%\vspace{-0.9cm}
%\end{figure}

\section{Conclusions and Discussion}
In this paper, we have studied a RIS-based downlink multi-user multi-antenna system in the absence of direct links between the BS and users. The BS transmits signals to users via the reflection-based RIS with limited discrete phase shifts. To better depict the close coupling between channel propagation and the RIS configuration pattern selection, we haved considered a reflection-dominated one-hop propagation model between the BS and users. Based on this model, we have carried out an HBF scheme for sum rate maximization where the continuous digital beamforming has been performed at the BS and the discrete analog beamforming has been achieved inherently at the RIS via configuration pattern selection. The sum rate maximization problem has been decomposed into two subproblems and solved iteratively by our proposed SRM algorithm.

Three remarks can be drawn from the theoretical analysis and numerical results, providing insights for RIS-based system design.
\begin{itemize}
	\item \emph{The sum rate of the RIS-based system with discrete phase shifts increases rapidly when the number of quantization bits $b$ is small, and gradually approaches the sum rate achieved in the continuous case if $b$ is large enough.}
	\item \emph{The sum rate increases with the size of RIS and converges to a stable value as the size of RIS grows to reach the threshold determined by Proposition~\ref{prop2}.}
	\item \emph{The minimum number of transmit antennas at the BS required to achieve any fully digital beamforming scheme is only half of that in traditional HBF schemes, implying that the RIS-based HBF scheme can greatly reduce the cost of dedicated hardware.}
\end{itemize}
The above remarks have indicated that when designing the RIS-based systems, a moderate size of RIS and a very small number of quantization bits are enough to achieve the satisfying sum rate at low cost.

%his implies that only a small number of quantization bits is required to achieve satisfying performance.

%we have proved that the minimum number of transmit antennas at the BS required to achieve any fully digital beamforming scheme is only half of that in traditional HBF schemes, implying that the RIS-based analog beamforming can greatly reduce the dedicated hardware required. Third, starting from the pure LoS case, we have derived new criteria for the RIS-based system design to obtain a high-rank channel matrix such that the maximum data rates can be achieved. Verified by the simulation results, we have explored how the size of RIS and the antenna array design at the BS influence the achievable rate given the placement of RIS. Specifically, there is an optimal RIS size at which the sum rate does not increase any more.

\vspace{-0.5cm}

\begin{appendices}
	
	\section{Proof of Proposition \ref{pro_3}}\label{proof_3}
	Note that
	\begin{align}
	%\begin{array}{ll}
	\tilde{\bm{F}}\tilde{\bm{F}}^H  &= \bm{P}^{-\frac{1}{2}}\sum\limits_{l_1,l_2}q_{l_1,l_2}(\bm{H}_{l_1,l_2} \circ \bm{\Phi}) \sum\limits_{{l_1}',{l_2}'}q^H_{{l_1}',{l_2}'}(\bm{H}_{{l_1}',{l_2}'} \circ \bm{\Phi})^H \bm{P}^{-\frac{1}{2}}\nonumber\\
	&= \sum\limits_{l_1,l_2}q_{l_1,l_2}\sum\limits_{{l_1}',{l_2}'}q^H_{{l_1}',{l_2}'} \bm{P}^{-\frac{1}{2}}(\bm{H}_{l_1,l_2} \circ \bm{\Phi})(\bm{H}_{{l_1}',{l_2}'} \circ \bm{\Phi})^H \bm{P}^{-\frac{1}{2}}\nonumber\\
	&= \sum\limits_{l_1,l_2}\sum\limits_{{l_1}',{l_2}'} \frac{(j + e^{j\theta_{l_1,l_2}})(-j + e^{-j\theta_{{l'}_1,{l'}_2}})}{4} \bm{A}_{l_1,l_2,{l_1}',{l_2}'}\nonumber\\
	&= \sum\limits_{l_1,l_2}\sum\limits_{{l_1}',{l_2}'} \frac{\bm{A}_{l_1,l_2,{l_1}',{l_2}'}}{4} \left(\cos(\theta_{l_1,l_2} - \theta_{{l_1}',{l_2}'}) + j \sin(\theta_{l_1,l_2} - \theta_{{l_1}',{l_2}'})\right) \nonumber\\
	&~~ + \sum\limits_{l_1,l_2}\sum\limits_{{l_1}',{l_2}'} \frac{\bm{A}_{l_1,l_2,{l_1}',{l_2}'}}{4} \left( \cos(\theta_{{l_1}',{l_2}'}) + \cos(\theta_{{l}_1,{l}_2}) + 1\right)\nonumber\\
	&= \sum\limits_{l_1,l_2}\sum\limits_{{l_1}',{l_2}'} \frac{\bm{A}_{l_1,l_2,{l_1}',{l_2}'}}{4} \left(\cos(\Delta \theta_{l_1,l_2,{l_1}',{l_2}'}) + j \sin(\Delta\theta_{l_1,l_2,{l_1}',{l_2}'})\right)\nonumber\\
	&~~ +\sum\limits_{l_1,l_2}\sum\limits_{{l_1}',{l_2}'} \frac{\bm{A}_{l_1,l_2,{l_1}',{l_2}'}}{4} \left( \cos(\theta_{{l_1}',{l_2}'}) + \cos(\theta_{{l}_1,{l}_2}) + 1\right),\nonumber\\
	%\end{array}
	\end{align}
	is not linear with respect to $\theta_{l_1,l_2}$. Taking the advantage of the discrete property of $\theta_{l_1,l_2}$, we can further transform the non-linear functions into linear ones.
	
	With the definitions of $\bm{x}^{l_1,l_2}$, we have
	\begin{equation}\label{transform}
	\cos(\theta_{l_1,l_2}) = \bm{x}^{l_1,l_2}\bm{c}^{T},~~\sin(\theta_{l_1,l_2}) = \bm{x}^{l_1,l_2}\bm{s}^{T},
	\end{equation}
	with $\|\bm{x}^{l_1,l_2}\|_1 = 1$. It is also worthwhile to point out that the value of $\theta_{l_1,l_2}$ only falls in the range $[0,2\pi)$, and thus, we have
	\begin{equation}
	\bm{e}^T\bm{x}^{l_1,l_2} = 0.
	\end{equation}
	
	Similarly, according to the definitions of $\bm{y}^{l_1,l_2,{l_1}',{l_2}'}$, we have
	\begin{equation}
	\cos(\Delta \theta_{l_1,l_2,{l_1}',{l_2}'}) = \bm{y}^{l_1,l_2,{l_1}',{l_2}'}\bm{c}^{T},~~\sin(\Delta \theta_{l_1,l_2,{l_1}',{l_2}'}) = \bm{y}^{l_1,l_2,{l_1}',{l_2}'}\bm{s}^{T},
	\end{equation}
	with
	\begin{equation}
	\bm{a}^{T}(\bm{x}^{l_1,l_2} - \bm{x}^{{l_1}',{l_2}'}) = \bm{a}^{T}\bm{y}^{l_1,l_2,{l_1}',{l_2}'}.
	\end{equation}
	
	With these transformations, $\tilde{\bm{F}}\tilde{\bm{F}}^H$ is linear with respective to $\bm{x}^{l_1,l_2}$ and $\bm{y}^{l_1,l_2,{l_1}',{l_2}'}$.
	
	\section{Proof of Proposition \ref{prop1}} \label{proof_prop1}
	We first prove that when $N_t = K$, any fully digital beamforming scheme can be achieved by the RIS-based HBF scheme if $N_R^2 \ge K{N_t}$ holds. We then state that it is not possible to achieve digital beamforming when $N_t < K$. Therefore, $N_t \ge 2K$ is a sufficient condition.
	
	i) Denote the channel matrix between the RIS and the users as ${H_{FD}} \in {^{K \times N_R^2}}$. Since $N_t = K$, the $k$th column of the digital beamformer can be expressed by ${\textbf{V}_{D,k}} = {\left[ {{0^T},{v_{k,k}},{0^T}} \right]^T}$, where $v_{k,k}$ is the $k$th element of the the $k$th column in ${\textbf{V}}_{D} \in {\mathbb{R}}^{N_t \times K}$. To satisfy ${{\textbf{H}}_{FD}}{{\textbf{V}}_{FD}} = {\textbf{F}}{{\textbf{V}}_D}$, we have
	\begin{equation}
	\left[ { \ldots ,{f_{m,k}}, \ldots } \right]\left[ {\begin{array}{*{20}{c}}
		0\\
		\vdots \\
		{{v_{k,k}}}\\
		\vdots \\
		0
		\end{array}} \right] = {\left( {{{\textbf{H}}_{FD}}{{\textbf{V}}_{FD}}} \right)_{m,k}},
	\end{equation}
	i.e.,
	\begin{equation} \label{FD1}
	\sum\limits_{{l_1},{l_2}} {\frac{1}{2}{\phi ^{\left( k \right)}}\left[ {\left( {j + {e^{j{\theta _{{l_1},{l_2}}}}}} \right)g_{{l_1},{l_2}}^{m,k}} \right]} {v_{k,k}} = {\left( {{H_{FD}}{V_{FD}}} \right)_{m,k}},
	\end{equation}
	for all $0 \le m,k \le K$, where $f_{m,k}$ is the element of matrix $\textbf{F}$. Note that the term multiplied by $v_{k,k}$ in $\left( \ref{FD1} \right)$ can achieve different magnitudes owning to the linear combination of channel coefficients, which is different from the traditional HBF~\cite{XAS-2005}. At least one solution can be found for this set of equations if the number of equations is no smaller than the number of variables, i.e., $N_R^2 \ge K{N_t}$. This also holds for $N_t > K$ since we can always use the solution for $N_t = K$ and set those extra variables to be zero.
	
	ii) We observe that $rank\left( {{{\textbf{H}}_{FD}}{{\textbf{V}}_{FD}}} \right) = K$ and $rank\left( {{\textbf{F}}{{\textbf{V}}_D}} \right) = \min \left\{ {K,{N_t}} \right\}$. If $N_t < K$, then $rank\left( {{{\textbf{H}}_{FD}}{{\textbf{V}}_{FD}}} \right) > rank\left( {{\textbf{F}}{{\textbf{V}}_D}} \right)$, implying that the RIS-based HBF cannot implement the fully digital beamforming scheme. This completes the proof.
	
	\section{Proof of Proposition \ref{prop2}} \label{proof_prop2}
	Note that each user $k$ goes through different path losses due to various positions, which naturally varies the corresponding channel coefficients. Therefore, we only focus on orthogonalizing different links from the BS to the same user. Since the path loss is the same for different links with respect to user $k$, we consider the channel response between one BS antenna $n$ and all RIS elements as
	\begin{equation} \label{LOS1}
	{{\textbf{f}}^{\left( {n,k} \right)}} = \left[ {{q_{0,0}}{e^{ - j\frac{{2\pi }}{\lambda }\left( {D_{0,0}^{\left( n \right)} + d_{0,0}^{\left( k \right)}} \right)}},{q_{0,1}}{e^{ - j\frac{{2\pi }}{\lambda }\left( {D_{0,1}^{\left( n \right)} + d_{0,1}^{\left( k \right)}} \right)}}, \cdots ,{q_{{N_R} - 1,{N_R} - 1}}{e^{ - j\frac{{2\pi }}{\lambda }\left( {D_{{N_R} - 1,{N_R} - 1}^{\left( n \right)} + d_{{N_R} - 1,{N_R} - 1}^{\left( k \right)}} \right)}}} \right].
	\end{equation}
	To keep any two different links orthogonal to each other, the following condition should be satisfied,
	\begin{equation} \label{LOS2}
	{\left[ {{{\textbf{f}}^{\left( {{n_a},k} \right)}}} \right]^H} \cdot {{\textbf{f}}^{\left( {{n_b},k} \right)}} = 0,\forall {n_a} \ne {n_b},
	\end{equation}
	where $n_a$ and $n_b$ denote two different transmit antennas.
	
	By substituting $\left( \ref{LOS1} \right)$, $\left( \ref{LOS2} \right)$ can be rewritten by
	\begin{equation}\label{LOS3}
	\sum\limits_{{l_2} = 0}^{{N_R} - 1} {\sum\limits_{{l_1} = 0}^{{N_R} - 1} {{q_{{l_1},{l_2}}} \cdot {{\left( {{q_{{l_1},{l_2}}}} \right)}^*}} } {e^{j\frac{{2\pi }}{\lambda }\left( {D_{{l_1},{l_2}}^{\left( {{n_a}} \right)} - D_{{l_1},{l_2}}^{\left( {{n_b}} \right)}} \right)}} = 0.
	\end{equation}
	We substitute $\left( \ref{q_value} \right)$ and $\left( \ref{distance_BS_RIS} \right)$ into $\left( \ref{LOS3} \right)$ and obtain the following
	\begin{equation} \label{LOS4}
	\sum\limits_{{l_1} = 0}^{{N_R} - 1} {\left[ {{e^{j\frac{{2\pi }}{{\lambda D_{0,0}^{\left( 0 \right)}}}{l_1}d_R^{\left( 1 \right)}{d_B}\left( {{n_b} - {n_a}} \right)\cos {\theta _R}\cos {\theta _B}}} \cdot \sum\limits_{{l_2} = 1}^{{N_R} - 1} {\left( {1 + \sin {\theta _{{l_1},{l_2}}}} \right)} } \right]}  = 0,
	\end{equation}
	where $n_b - n_a \in \mathbb{Z}$. Since this condition should hold for any two transmit antennas $n_a$ and $n_b$, we then have $\left( \ref{RIS_design} \right)$ based on the principle of geometric sums.
	
%\section{Proof of Proposition \ref{pro_sym}}\label{proof_sym}
%First, it is easy to check that $\tilde{\bm{F}}\tilde{\bm{F}}^H$ is a symmetric matrix, because $(\tilde{\bm{F}}\tilde{\bm{F}}^H)^H = \tilde{\bm{F}}\tilde{\bm{F}}^H$.
%
%Then, let $\bm{v}$ be an arbitrary vector, and we have
%\begin{equation}
%\bm{v}^{H}\tilde{\bm{F}}\tilde{\bm{F}}^H\bm{v} = (\bm{v}^{H}\tilde{\bm{F}})(\bm{v}^{H}\tilde{\bm{F}})^H \geq 0.
%\end{equation}
%Therefore, $\tilde{\bm{F}}\tilde{\bm{F}}^H$ is also a positive semi-definite matrix.
	
\end{appendices}

\end{document}